\documentclass[nohyperref]{article}

\usepackage{microtype}
\usepackage{graphicx}
\usepackage{booktabs} %

\usepackage{hyperref}

\usepackage[accepted]{icml2022}

\usepackage{amsmath}
\usepackage{amssymb}
\usepackage{mathtools}
\usepackage{amsthm}

\usepackage[capitalize,noabbrev]{cleveref}

\theoremstyle{plain}

\theoremstyle{definition}

\theoremstyle{remark}

\usepackage{courier}
\usepackage{caption}
\usepackage{comment}
\usepackage{color}
\usepackage{bm}
\usepackage{xspace}
\usepackage{enumitem}
\usepackage{multirow}
\usepackage[bottom]{footmisc}
\usepackage{subcaption}
\usepackage{wrapfig}
\usepackage{soul}
\usepackage{amsthm}
\usepackage{nicefrac}       %
\usepackage{amsbsy}
\usepackage{bbm}
\usepackage{stfloats}
\usepackage{mathrsfs}
\usepackage{thmtools}
\usepackage{thm-restate}
\usepackage{xr}
\usepackage{tabularx}

\def\Ex{\mathop{\mathbb{E}}}
\DeclareMathOperator{\E}{\mathbb{E}}
\DeclareMathOperator{\R}{\mathbb{R}}
\DeclarePairedDelimiter\floor{\lfloor}{\rfloor}
\newcommand{\norm}[1]{\left\lVert#1\right\rVert}

\newcommand{\inner}[1]{\left\langle#1\right\rangle}

\DeclareMathOperator*{\argmax}{arg\,max}

\newcommand{\ber}[1]{\mathrm{Bern}\left(#1\right)}

\def\eqref#1{Eqn.~(\ref{#1})}
\def\figref#1{Fig.~\ref{#1}}

\newcommand{\rt}{RT\xspace}
\newcommand{\art}{AdvRT\xspace}

\makeatletter
\DeclareRobustCommand\onedot{\futurelet\@let@token\@onedot}
\def\@onedot{\ifx\@let@token.\else.\null\fi\xspace}

\icmltitlerunning{Demystifying the Adversarial Robustness of Random Transformation Defenses}

\begin{document}

\twocolumn[
    \icmltitle{Demystifying the Adversarial Robustness of Random Transformation Defenses}

    \icmlsetsymbol{equal}{*}

    \begin{icmlauthorlist}
        \icmlauthor{Chawin Sitawarin}{ucb}
        \icmlauthor{Zachary Golan-Strieb}{ucb}
        \icmlauthor{David Wagner}{ucb}
    \end{icmlauthorlist}

    \icmlaffiliation{ucb}{Department of Electrical Engineering and Computer Sciences, University of California, Berkeley, Berkeley CA, USA}

    \icmlcorrespondingauthor{Chawin Sitawarin}{chawins@berkeley.edu}

    \icmlkeywords{Machine Learning, ICML, Adversarial Examples, Robustness, Computer Vision}

    \vskip 0.3in
]

\printAffiliationsAndNotice{}  %

\begin{abstract}
    Neural networks' lack of robustness against attacks raises concerns in security-sensitive settings such as autonomous vehicles.
    While many countermeasures may look promising, only a few withstand rigorous evaluation.
    Defenses using random transformations (\rt) have shown impressive results, particularly BaRT~\citep{raff_barrage_2019} on ImageNet.
    However, this type of defense has not been rigorously evaluated, leaving its robustness properties poorly understood.
    Their stochastic properties make evaluation more challenging and render many proposed attacks on deterministic models inapplicable.
    First, we show that the BPDA attack~\citep{athalye_obfuscated_2018} used in BaRT's evaluation is ineffective and likely overestimates its robustness.
    We then attempt to construct the strongest possible \rt defense through the informed selection of transformations and Bayesian optimization for tuning their parameters.
    Furthermore, we create the strongest possible attack to evaluate our \rt defense.
    Our new attack vastly outperforms the baseline, reducing the accuracy by 83\% compared to the 19\% reduction by the commonly used EoT attack ($4.3\times$ improvement).
    Our result indicates that the \rt defense on Imagenette dataset (a ten-class subset of ImageNet) is not robust against adversarial examples.
    Extending the study further, we use our new attack to adversarially train \rt defense (called \art), resulting in a large robustness gain.
    Code is available at \href{https://github.com/wagner-group/demystify-random-transform}{https://github.com/wagner-group/demystify-random-transform}.
\end{abstract}

\section{Introduction} \label{sec:introduction}

Today, deep neural networks are widely deployed in safety-critical settings such as autonomous driving and cybersecurity.
Despite their effectiveness at solving a wide-range of challenging problems, they are known to have a major vulnerability. Tiny crafted perturbations added to inputs (so called \emph{adversarial examples}) can arbitrarily manipulate the outputs of these large models, posing a threat to the safety and privacy of the millions of people who rely on existing ML systems. 
The importance of this problem has drawn substantial attention, and yet the research community has not devised a concrete countermeasure.

Adversarial training~\citep{madry_deep_2018} has been the foremost approach for defending against adversarial examples.
While adversarial training provides increased robustness, it results in a loss of accuracy on benign inputs. 
Recently, a promising line of defenses against adversarial examples has emerged. 
These defenses randomize either the model parameters or the inputs themselves~\citep{lecuyer_certified_2019,he_parametric_2019,liu_advbnn_2019,xie_mitigating_2018,zhang_defending_2019,bender_defense_2020,liu_robust_2018,cohen_certified_2019,dhillon_stochastic_2018}. 
Introducing randomness into the model can be thought of as a form of smoothing that removes sinuous portions of the decision boundary where adversarial examples frequently lie~\citep{he_decision_2018}. 
Other works attribute its success to the ensemble~\citep{guo_countering_2018} or the ``moving-target''~\citep{chen_evaluating_2021} effect.
Among these randomization approaches, \citet{raff_barrage_2019} propose Barrage of Random Transforms (BaRT), a new defense which applies a large set of random image transformations to classifier inputs. 
They report a $24\times$ increase in robust accuracy over previously proposed defenses.

Despite these promising results, researchers still lack a clear understanding of how to properly evaluate random defenses. 
This is concerning as a defense can falsely appear more robust than it actually is when evaluated using sub-optimal attacks~\citep{athalye_obfuscated_2018,tramer_adaptive_2020}.
Therefore, in this work, we improve existing attacks on randomized defenses, and use them to rigorously evaluate BaRT and more generally, random transformation (\rt) defenses.
We find that sub-optimal attacks have led to an overly optimistic view of these \rt defenses.
Notably, we show that even our best \rt defense is much less secure than previously thought, formulating a new attack that reduces its security (from 70\% adversarial accuracy found by the baseline attack to only 6\% on Imagenette).

We also take the investigation further and combine \rt defense with adversarial training. 
Nevertheless, this turns out to be ineffective as the attack is not sufficiently strong and only generates weak adversarial examples for the model to train with.
The outcomes appear more promising for CIFAR-10, but it still lacks behind deterministic defense such as \citet{madry_deep_2018} and \citet{zhang_theoretically_2019}.
We believe that stronger and more efficient attacks on \rt-based models will be necessary not only for accurate evaluation of the stochastic defenses but also for improving the effectiveness of adversarial training for such models.

To summarize, we make the following contributions:
\begin{itemize}[noitemsep]
    \item We show that non-differentiable transforms impede optimization during an attack and even an adaptive technique for circumventing non-differentiability (i.e., BPDA~\citep{athalye_obfuscated_2018}) is not sufficiently effective. This reveals that existing \rt defenses are likely non-robust.
    \item To this end, we suggest that an \rt defense should only use differentiable transformations for reliable evaluations and compatibility with adversarial training.
    \item We propose a new state-of-the-art attack for \rt defense that improves over EoT~\citep{athalye_synthesizing_2018} in terms of both the loss function and the optimizer. We explain the success of our attack through the variance of the gradients.
    \item Improve the \rt scheme by using Bayesian optimization for hyperparameter tuning and combining it with adversarial training which uses our new attack method instead of the baseline EoT. 
\end{itemize}

\section{Background and Related Works} \label{sec:background}

\subsection{Adversarial Examples}

Adversarial examples are carefully perturbed inputs designed to fool a machine learning model~\cite{szegedy_intriguing_2014,biggio_evasion_2013,goodfellow_explaining_2015}. 
An adversarial perturbation $\delta$ is typically constrained to be within some $\ell_p$-norm ball with a radius of $\epsilon$.
The $\ell_p$-norm ball is a proxy to the ``imperceptibility'' of $\delta$ and can be thought of as the adversary's budget.
In this work, we primarily use $p = \infty$ and only consider adaptive white-box adversary.
Finding the worst-case perturbation $\delta^*$ requires solving the following optimization problem:
\begin{align} \label{eq:adv}
     x_{\text{adv}} = x + \delta^* = x + \argmax_{\delta : \norm{\delta}_p \le \epsilon} ~L(x + \delta, y)
\end{align}
where $L:\mathbb{R}^d \times \mathbb{R}^C \to \mathbb{R}$ is the loss function of the target model which, in our case, is a classifier which makes predictions among $C$ classes. 
Projected gradient descent (PGD) is often used to solve the optimization problem in \eqref{eq:adv}.

\subsection{Randomization Defenses}

A number of recent papers have proposed defenses against adversarial examples which utilize inference-time randomization. 
One common approach is to sample weights of the network from some probability distribution~\citep{liu_robust_2018,he_parametric_2019,liu_advbnn_2019,bender_defense_2020}. 
In this paper, we instead focus on defenses that apply random transforms to the input~\citep{raff_barrage_2019,xie_mitigating_2018,zhang_defending_2019,cohen_certified_2019}, many of which claim to achieve state-of-the-art robustness. 
Unlike prior evaluations, we test these defenses using a wide range of white-box attacks as well as a novel stronger attack.
A key issue when evaluating these schemes is that PGD attacks require gradients through the entire model pipeline, but many defenses use non-differentiable transforms.
As we show later, this can cause evaluation results to be misleading.

Various random transformation defenses have been proposed.
\citet{xie_mitigating_2018} randomly resize and pad the images. 
While this defense ranked second in the NeurIPS 2017 adversarial robustness competition, they did not consider in their evaluation adaptive attacks where the adversary has full knowledge of the transformations.
\citet{zhang_defending_2019} add Gaussian noise to the input and then quantize it. 
Their defense is reported to outperform all of the NeurIPS 2017 submissions. 
The adaptive attack used to evaluate their defense approximates the gradient of the transformations which could lead to a sub-optimal attack. 
In this paper, we use the exact gradients for all transforms when available.

More recently, \citet{raff_barrage_2019} claims to achieve a state-of-the-art robust accuracy $24\times$ better than adversarial training using a random transformation defense known as Barrage of Random Transforms (BaRT). 
BaRT involves randomly sampling a large set of image transformations and applying them to the input in random order. 
Because many transformations are non-differentiable, BaRT evaluates their scheme using PGD attack that approximates the gradients of the transformations. 
In Section~\ref{sec:bpda}, we show that this approximation is ineffective, giving overly optimistic impression of BaRT's robustness, and we re-evaluate BaRT using a stronger attack which utilizes exact transform gradients.

\begin{figure}[t!]
    \centering
    \includegraphics[width=0.5\textwidth]{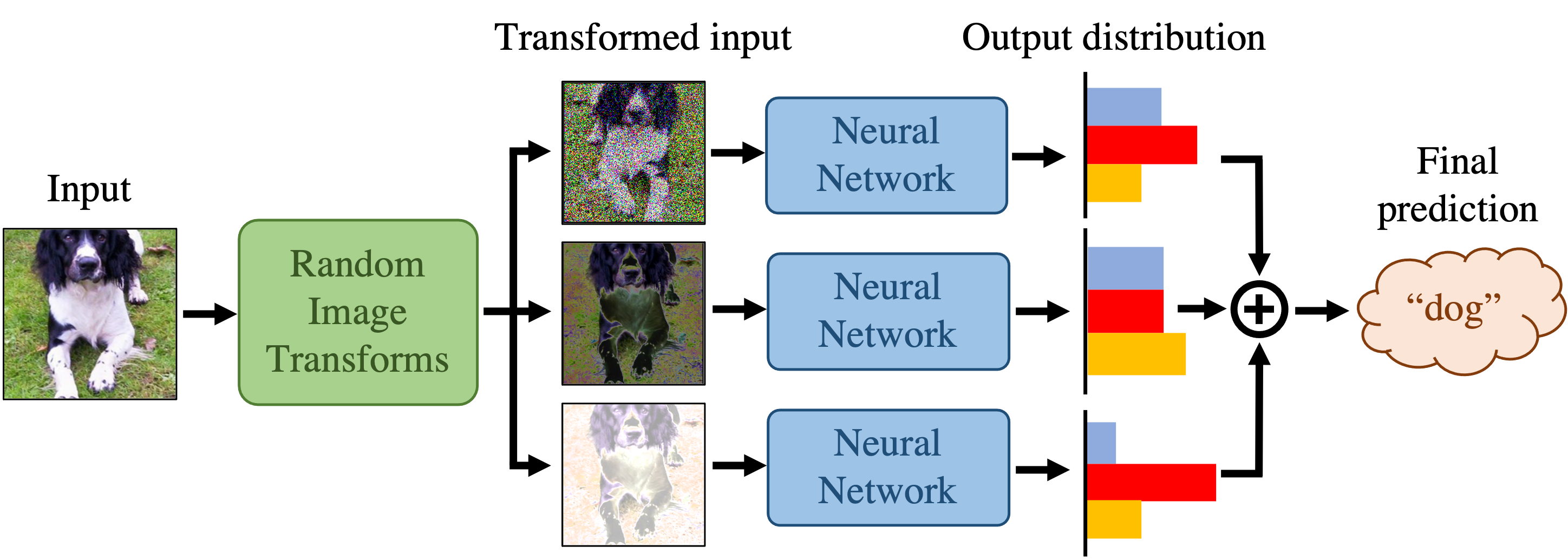}
    \caption{An illustration of a random transformation (\rt) defense against adversarial examples. Transformations of different types and parameters are sampled and applied sequentially to multiple copies of the input. All of the transformed inputs are then passed to a single neural network, and the outputs are combined to make the final prediction.}
    \label{fig:rt_diagram}
\end{figure}

\section{Random Transformation Defense} \label{ssec:random_transform}

Here, we introduce notations and the design of our \rt defense, formalizing the BaRT defense.

\subsection{Decision Rules} \label{sssec:rt}

\rt repeatedly applies a randomly chosen transform to the input, uses a neural network to make a prediction, and then averages the softmax prediction scores:
\begin{align} \label{eq:rt}
    g(x) \coloneqq \E_{\theta \sim p(\theta)} \left[ \sigma \left( f \left( t(x;\theta) \right) \right) \right]
\end{align}
where $\sigma(\cdot)$ is the softmax function, $f:\R^d\to\R^C$ a neural network ($C$ is the number of classes), and the transformation $t(\cdot;\theta):\R^d \to \R^d$  is parameterized by a random variable $\theta$ drawn from some distribution $p(\theta)$.

In practice, we approximate the expectation in \eqref{eq:rt} with $n$ Monte Carlo samples per one input $x$:
\begin{align} \label{eq:rt-approx}
    g(x) \approx g_n(x) \coloneqq \frac{1}{n} \sum_{i=1}^n \sigma\left( f(t(x;\theta_i)) \right)
\end{align}
We then define the final prediction as the class with the largest softmax probability: $\hat{y}(x) = \argmax_{c \in [C]}~[g_n(x)]_c$. 
Note that this decision rule is different from most previous works that use a majority vote on hard labels, i.e., $\hat{y}_{\mathrm{maj}}(x) = \argmax_{c \in [C]}~\sum_{i=1}^n \mathbbm{1}\left\{c = \argmax_{j \in [C]}~f_j(x)\right\}$~\cite{raff_barrage_2019,cohen_certified_2019}.
We later show in Appendix~\ref{ap:ssec:rule} that our rule is empirically superior to the majority vote.
From the Law of Large Numbers, as $n$ increases, the approximation in \eqref{eq:rt-approx} converges to the expectation in \eqref{eq:rt}.
\figref{fig:rt_diagram} illustrates the structure and the components of the \rt architecture.

\subsection{Parameterization of Transformations} \label{ssec:tf_params}

Here, $t(\cdot;\theta)$ represents a composition of $S$ different image transformations where $\theta = \{\theta^{(1)},\dots,\theta^{(S)}\}$ and $\theta^{(s)}$ denotes the parameters for the $s$-th transformation, i.e.,
\begin{align}
    t(x;\theta) = t_{\theta^{(S)}} \circ t_{\theta^{(S-1)}} \circ \dots \circ t_{\theta^{(1)}}(x)
\end{align}
Each $\theta^{(s)}$ is a random variable comprised of three components, i.e., $\theta^{(s)}=\{\tau^{(s)},\beta^{(s)},\alpha^{(s)}\}$, which dictate the properties of a transformation: 
\begin{enumerate}[noitemsep]
    \item \emph{Type} $\tau$ of transformation to apply (e.g., rotation, JPEG compression), which is uniformly drawn, without replacement, from a pool of $K$ transformation types: $\tau \sim \text{Cat}(K, \bm{1}/K)$.
    \item A \emph{boolean} $\beta$ indicating whether the transformation will be applied. This is a Bernoulli random variable with probability $p_\beta$: $\beta \sim \ber{p}$.
    \item \emph{Strength} of the transformation (e.g., rotation angle, JPEG quality) denoted by $\alpha$, sampled from a predefined distribution (either uniform or normal): $\alpha \sim p(a)$.
\end{enumerate}
Specifically, for each of the $n$ transformed samples, we sample a permutation of size $S$ out of $K$ transformation types in total, i.e. $\{\tau^{(1)},\dots,\tau^{(S)}\} \in \mathrm{Perm}(K, S)$.
Then the boolean and the strength of the $s$-th transform are sampled: $\beta^{(s)} \sim \ber{p_{\tau^{(s)}}}$ and $\alpha^{(s)} \sim p(a_{\tau^{(s)}})$.
We abbreviate this sampling process as $\theta \sim p(\theta)$ which is repeated for every transformed sample (out of $n$) for a single input. 

Assuming that the $K$ transformation types are fixed, an \rt defense introduces, at most, $2K$ hyperparameters, $\{p_1,\dots,p_K\}$ and $\{a_1,\dots,a_K\}$, that can be tuned.
It is also possible to tune by selecting $K'$ out of $K$ transformation types, but this is combinatorially large in $K$. 
In Appendix~\ref{ap:sec:bayes}, we show a heuristic for ``pruning'' the transformation types through tuning $p$ and $a$ (e.g., setting $p=0$ is equivalent to removing that transformation type).

\subsection{Choices of Transformations} \label{sssec:tf}

In this work, we use a pool of $K=33$ different image transformations including 19 differentiable and 2 non-differentiable transforms taken from the 30 BaRT transforms~\cite{raff_barrage_2019} (counting each type of noise injection as its own transform). 
We replace non-differentiable transformations with a smooth differentiable alternative~\cite{shin_jpegresistant_2017}.
The transformations fall into seven groups: noise injection (7), blur filtering (4), color-space alteration (8), edge detection (2), lossy compression (3), geometric transformation (5), and stylization (4).
All transforms are described in Appendix~\ref{ap:ssec:tf_list}.
\section{Evaluating \citet{raff_barrage_2019}'s BaRT} \label{sec:bpda}

Backward-pass differentiable approximation (BPDA) was proposed as a heuristic for approximating gradients of non-differentiable components in many defenses to make gradient-based attacks applicable~\citep{athalye_obfuscated_2018}.
It works by first approximating the function with a neural network and backpropagate through this network instead of the non-differentiable function.  
Evaluations of BaRT in \citet{raff_barrage_2019} have considered BPDA as some transformations are innately non-differentiable or have zero gradients almost everywhere (e.g., JPEG compression, precision reduction, etc.). 
To approximate a transformation, we train a model $\tilde{t}_\phi$ that minimizes the Euclidean distance between the transformed image and the model output:
\begin{align} \label{eq:bpda_loss}
    \min_{\phi}~\sum_{i=1}^N\Ex_{\theta \sim p(\theta)}\norm{\tilde{t}_\phi(x_i; \theta) - t(x_i; \theta)}_2
\end{align}
We evaluate the BPDA approximation below in a series of experiments that compare the effectiveness of the BPDA attack to an attack that uses exact gradients.

\subsection{Experiment Setup}

Our experiments use two datasets: CIFAR-10 and Imagenette~\citep{howard_fastai_2021}, a ten-class subset of ImageNet.
While CIFAR-10 is the most common benchmark in the adversarial robustness domain, some image transformations work poorly on low-resolution images.
We choose Imagenette because BaRT was created on ImageNet, but we do not have resources to do thorough investigation on top of adversarial training on ImageNet. 
Additionally, the large and realistic images from Imagenette more closely resemble real-world usage
All Imagenette models are pre-trained on ImageNet to speed up training and boost  performance.
Since \rt models are stochastic, we report their average accuracy together with the 95\% confidence interval from 10 independent runs.
Throughout this work, we consider the perturbation size $\epsilon$ of $16/255$ for Imagenette and $8/255$ for CIFAR-10.
Appendix~\ref{ap:ssec:exp_setup} has more details on the experiments (network architecture, hyperparameters, etc.).

\subsection{BPDA Attack is Not Sufficiently Strong} \label{ssec:bpda-exp}

\begin{table*}[t]
    \small
    \centering
    \caption{Comparison of attacks with different gradient approximations. ``Exact'' directly uses the exact gradient. ``BPDA'' uses the BPDA gradient for most transforms and the identity for a few. ``Identity'' backpropagates as an identity function, and ``Combo'' uses exact gradient for differentiable transforms and BPDA gradient otherwise. Full BaRT uses a nearly complete set of BaRT transforms ($K=26$), and ``BaRT (only differentiable)'' uses only differentiable transforms ($K = 21$). We use PGD attack with EoT and CE loss ($\epsilon = 16/255$, 40 steps).}
    \label{tab:bpda}
    \begin{tabular}{lrrrrr}
    \toprule
    \multirow{2}{*}{Transforms used} & \multirow{2}{*}{Clean accuracy} & \multicolumn{4}{c}{Adversarial accuracy w/ gradient approximations} \\ 
    \cmidrule(l){3-6}
    &  & Exact & BPDA & Identity & Combo \\ \midrule
    BaRT (full) & $88.10 \pm 0.16$ & n/a & $52.32 \pm 0.22$ & $36.49 \pm 0.25$ & $\mathbf{25.24 \pm 0.16}$ \\
    BaRT (only differentiable) & $87.43 \pm 0.28$ & $\mathbf{26.06 \pm 0.21}$ & $65.28 \pm 0.25$ & $41.25 \pm 0.26$ & n/a \\
    \bottomrule
    \end{tabular}
\end{table*}

\begin{figure}
    \centering
    \begin{subfigure}[b]{0.3\linewidth}
        \centering
        \includegraphics[width=\linewidth]{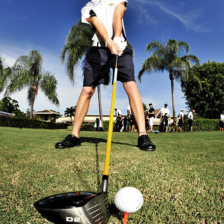}
        \caption{Original}
    \end{subfigure}
    \hfill
    \begin{subfigure}[b]{0.3\linewidth}
        \centering
        \includegraphics[width=\linewidth]{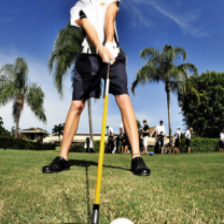}
        \caption{Exact crop}
    \end{subfigure}
    \hfill
    \begin{subfigure}[b]{0.3\linewidth}
        \centering
        \includegraphics[width=\linewidth]{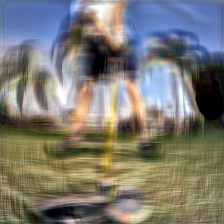}
        \caption{BPDA crop}
    \end{subfigure}
    \caption{Comparison of crop transform output and output of BPDA network trained to approximate crop transform.}
    \label{fig:zoom_comparison}
\end{figure}

We re-implemented and trained a BaRT model on these datasets, and then evaluated the effectiveness of BPDA attacks against this model.\footnote{The authors have been very helpful with the implementation details but cannot make the official code or model weights public.}
First, we evaluate the full BaRT model in Table~\ref{tab:bpda}, comparing an attack that uses a BPDA approximation (as \citet{raff_barrage_2019}) vs an attack that uses the exact gradient for differentiable transforms and BPDA for non-differentiable transforms, denoted ``BPDA'' and ``Combo'', respectively.
Empirically, we observe that attacks using BPDA are far weaker than the equivalent attack using exact gradient approximations.
Similarly, on a variant BaRT model that uses only the subset of differentiable transforms, the BDPA attack is worse than an attack that uses the exact gradient for all transforms.
BPDA is surprisingly weaker than even a naive attack which approximates all transform gradients with the identity.
There are a few possible explanations for the inability of BPDA to approximate transformation gradients well:
\begin{enumerate}[noitemsep]
    \item As \figref{fig:zoom_comparison} illustrates, BPDA struggles to approximate some transforms accurately.
    This might be partly because the architecture \citet{raff_barrage_2019} used (and we use) to approximate each transform has limited functional expressivity:
    it consists of five convolutional layers with 5x5 kernel and one with 3x3 kernel (all strides are 1), so a single output pixel can only depend on the input pixels fewer than 11 spaces away in any direction ($5 \cdot \floor{\frac{5}{2}} + 1 \cdot \floor{\frac{3}{2}} = 11$). 
    Considering the inputs for Imagenette are of size $224\times 224$, some transforms like ``crop'' which require moving pixels much longer distances are impossible to approximate with such an architecture.
    \item The BPDA network training process for solving \eqref{eq:bpda_loss} may only find a sub-optimal solution, yielding a poor approximation of the true transformation.
    \item During the attack, the trained BPDA networks are given partially transformed images, yet the BPDA networks are only trained with untransformed inputs.
    \item Since we are backpropagating through several transforms, one poor transform gradient approximation could ruin the overall gradient approximation. 
\end{enumerate}
Appendix \ref{ap:ssec:bpda_detail} has more details on these experiments.
These results show that BaRT's evaluation using BPDA was overly optimistic, and BaRT is not as robust as previously thought. 

Since BPDA is unreliable for approximating gradients of non-differentiable image transformations, \textbf{we recommend that other ensuing \rt-based defenses only use differentiable transformations.}
For the rest of this paper, we only study the robustness of \rt defenses with differentiable transforms to isolate them from an orthogonal line of research on non-differentiable defenses (e.g., with approximate gradients or zero-th order attacks).
Additionally, differentiable models can boost their robustness further when combined with adversarial training.
We explore this direction in Section~\ref{sec:combine_at}.
Even without non-differentiable transforms, we still lack reliable evaluation on stochastic defenses apart from EoT.
In the next section, we show that applying an EoT attack on \rt defense results in a critically sub-optimal evaluation.
After that, we propose a stronger attack.

\section{Hyperparameter Tuning on \rt Defenses} \label{sec:bayesopt}

Before investigating attacks, we want to ensure we evaluate on the most robust \rt defense possible.
We found that BaRT is not robust, but it could be because of the chosen transformations and their hyperparameters which they do not provide any justification for.
Finding the most robust \rt defense is, however, challenging because it consists of numerous hyperparameters including the $K$ transformation types, the number of transformations to apply ($S$), and their parameters ($a$ and $p$).
A typical grid search is intractable since we have 33 transformations, and trying to optimize the parameters directly with the reparameterization trick does not work as most transforms are not differentiable w.r.t. their parameters. 

We systematically address this problem by using Bayesian optimization (BO)~\cite{snoek_practical_2012}, a well-known black-box optimization technique used for hyperparameter search, to fine-tune $a$ and $p$.
In short, BO optimizes an objective function that takes in the hyperparameters ($a$ and $p$ in our case) as inputs and outputs adversarial accuracy.
This process, which is equivalent to one iteration in BO, is computationally expensive as it involves training a neural network as a backbone for an \rt defense and evaluating it with our new attack.
Consequently, we have to scale down the problem by shortening the training, using fewer training/testing data samples, and evaluating with fewer attack steps.
Essentially, we have to trade off precision of the search for efficiency.
Because BO does not natively support categorical or integral variables, we experiment with different choices for $K$ and $S$ without the use of BO.
The full details of this procedure are presented Appendix~\ref{ap:sec:bayes}.
\section{State-of-the-Art Attack on \rt Defenses} \label{sec:attack}

\begin{table}[t!]
\small
\centering
\caption{Comparison between the baseline EoT attack~\citep{athalye_synthesizing_2018}, AutoAttack~\citep{croce_reliable_2020}, and our attack on the \rt defense whose transformation parameters have been fine-tuned by Bayesian Optimization to maximize the robustness. 
For AutoAttack, we use its standard version combined with EoT.
For Imagenette, we use $\epsilon=16/255$, for CIFAR-10, $\epsilon=8/255$. }
\label{tab:attack_compare}
\begin{tabular}{@{}lrr@{}}
\toprule
\multirow{2}{*}{Attacks} & \multicolumn{2}{c}{Accuracy} \\ \cmidrule{2-3}
 & CIFAR-10 & Imagenette \\ \midrule
No attack & $81.12 \pm 0.54$ & $89.04 \pm 0.34$  \\
Baseline & $33.83 \pm 0.44$ & $70.79 \pm 0.53$  \\
AutoAttack & $61.13 \pm 0.85$ & $85.46 \pm 0.43$ \\
Our attack & $\bm{29.91} \pm 0.35$ & $\bm{6.34} \pm 0.35$ \\
\bottomrule
\end{tabular}
\vspace{-10pt}
\end{table}

\begin{algorithm}[tb]
\caption{Our best attack on \rt defenses}
\label{alg:attack}
\begin{algorithmic}
    \STATE {\bf Input:} Set of $K$ transformations and distributions of their parameters $p(\theta)$, neural network $f$, perturbation size $\epsilon$, max. PGD steps $T$, step size $\{\gamma_t\}_{t=1}^T$, and AggMo's damping constants $\{\mu_b\}_{b=1}^B$.
    \STATE {\bfseries Output:} Adversarial examples $x_{\mathrm{adv}}$
    \STATE {\bfseries Data:} Test input $x$ and its ground-truth label $y$
    \STATE \textcolor{blue}{\texttt{// Initialize x\_adv and velocities}}
    \STATE $x_{\mathrm{adv}} \gets x + u \sim \mathcal{U}[-\epsilon,\epsilon],\quad \{v_b\}_{b=1}^B \gets \bm{0}$
    \STATE $x_{\mathrm{adv}} \gets \mathrm{Clip}(x_{\mathrm{adv}}, 0, 1)$
    \FOR{$t=1$ {\bfseries to} $T$}
        \STATE $\{\theta_i\}_{i=1}^n \sim p(\theta)$
        \STATE \textcolor{blue}{\texttt{// Compute a gradient estimate with linear loss on logits (Section~\ref{ssec:adv_obj}) and with SGM (Section~\ref{ssec:ensemble})}}
        \STATE $G_n \gets \nabla \mathcal{L}_{\mathrm{Linear}}\left(\frac{1}{n} \sum_{i=1}^n f(t(x_{\mathrm{adv}};\theta_i)), y\right)$
        \STATE $\hat{G}_n \gets \mathrm{sign}(G_n)$ \hfill \textcolor{blue}{\texttt{// Use signed gradients}}
        \STATE \textcolor{blue}{\texttt{Update velocities and x\_adv with AggMo (Section~\ref{ssec:optimizer})}}
        \FOR{$b=1$ {\bfseries to} $B$}
            \STATE $v_b \gets \mu_b \cdot v_b + \hat{G}_n$
        \ENDFOR
        \STATE $x_{\mathrm{adv}} \gets x_{\mathrm{adv}} + \frac{\gamma_t}{B}\sum_{b=1}^B v_b$
    \ENDFOR
\end{algorithmic}
\end{algorithm}

We propose a new attack on differentiable \rt defenses that leverages insights from previous literature on transfer attacks as well as recent stochastic optimization algorithms.
Our attack is immensely successful and shows that even the fine-tuned \rt defense from Section~\ref{sec:bayesopt} shows almost no adversarial robustness (Table~\ref{tab:attack_compare}).
We summarize our attack in Algorithm~\ref{alg:attack} before describing the setup and investigating the three main design choices that make this attack successful and outperform the baseline from \citet{athalye_synthesizing_2018} by a large margin.

\subsection{Setup: Stochastic Gradient Method} \label{ssec:var_sgd}

First, we describe the setup and explain intuitions around variance of the gradient estimates.
Finding adversarial examples on \rt defenses can be formulated as the following stochastic optimization problem:
\begin{align}
    \max_{\delta:\norm{\delta}_\infty \le \epsilon} H(\delta) &\coloneqq \max_{\delta:\norm{\delta}_\infty \le \epsilon} \E_{\theta} \left[h(\delta;\theta)\right] \\
    &\coloneqq \max_{\delta:\norm{\delta}_\infty \le \epsilon} \E_{\theta} \left[\mathcal{L}(f(t(x+\delta; \theta)), y)\right] \label{eq:sgd_setup}
\end{align}
for some objective function $\mathcal{L}$.
Note that we drop dependence on $(x,y)$ to declutter the notation.
Since it is not possible to evaluate the expectation or its gradients exactly, the gradients are estimated by sampling $\{\theta_i\}_{i=1}^n$ similarly to how we obtain a prediction $g_n$.
Suppose that $H$ is smooth and convex, and variance of the gradient estimates is bounded by $\sigma^2$, i.e.,
\begin{align} \label{eq:var}
    \Ex_{\theta \sim p(\theta)} \left[ \norm{\nabla h(\delta; \theta) - \nabla H(\delta)}^2 \right] \le \sigma^2,
\end{align}
the error of SGD after $T$ iterations is $\mathcal{O}\left(1/T + \sigma/\sqrt{T}\right)$ for an appropriate step size~\citep{ghadimi_stochastic_2013}.
This result suggests that small $\sigma$ or low-variance gradient speeds up convergence which is highly desirable for attackers and defenders alike.
Specifically, it leads to more efficient and more accurate evaluation as well as a stronger attack to use during adversarial training, which in turn, could yield a better defense (we explore this in Section~\ref{sec:combine_at}).

As a result, the analyses on our attack will be largely based on variance and two other measures of spread of the gradients.
Specifically, we measure (1) the dimension-averaged variance in \eqref{eq:var}, (2) cosine similarity and (3) a percentage of matching signs between mean gradient and each gradient sample.
Since all three metrics appear to be highly correlated in theory and in practice, we only report the variance in the main paper.
For the other metrics and their mathematical definitions, please see Appendix~\ref{ap:ssec:grad_var}.

\paragraph{EoT Baseline.}
We compare our attack to the baseline which is exactly taken from \citet{athalye_synthesizing_2018}.
This attack takes on the same form as \eqref{eq:sgd_setup} and its gradients are averaged over $n$ gradient samples:
\begin{align}
    H^{\mathrm{EoT}}_n(\delta) &\coloneqq \frac{1}{n} \sum_{j=1}^n~ \mathcal{L}\left(  f \left( t(x + \delta; \theta_j) \right), y\right) \label{eq:attack_eot}
\end{align}
It is important to note that this approximation does not exactly match the decision rule of \rt defenses as the expectation should be in front of $f$ but behind the loss function (see \eqref{eq:rt}). 
While the gradient estimates from \eqref{eq:attack_eot} are unbiased, they may have high variance as each gradient sample is equivalent to computing the loss on $g_n$ with $n=1$.
In the next section, we will compare other options for objective functions and decision rules and show that there are better alternatives to the original EoT.

\paragraph{Signed gradients.}
All of the attacks used in this study including ours and the baseline use signs of gradients instead of the gradients themselves.
This is a common practice for gradient-based $\ell_\infty$-attacks, and we have also empirically confirm that it leads to much stronger attacks.
This is also the reason that we measure sign matching as a measure of spread of the gradient estimates.
In addition to the $\ell_\infty$-constraint, using signed gradients as well as signed momentum is also beneficial as it has been shown to reduce variance for neural network training and achieve even faster convergence than normal SGD in certain cases~\citep{bernstein_signsgd_2018}.

\subsection{Adversarial Objectives and Decision Rules} \label{ssec:adv_obj}

Here, we propose new decision rules and loss functions for the attacks as alternatives to EoT.
Note that this need not be the same as the rule used for making prediction in \eqref{eq:rt}.
First, we introduce \emph{softmax} and \emph{logits} rules: 
\begin{align} 
&H^{\mathrm{softmax}}(\delta) \coloneqq \mathcal{L}\left(  \Ex_{\theta\sim p(\theta)} \left[ \sigma \left( f \left( t(x + \delta; \theta) \right) \right) \right], y\right) \\
&H^{\mathrm{logits}}(\delta) \coloneqq \mathcal{L} \left( \Ex_{\theta\sim p(\theta)} \left[ f \left( t(x + \delta; \theta) \right) \right], y\right) \label{eq:attack_logits}
\end{align}
$H^{\mathrm{softmax}}$, or loss of the expected softmax probability, is the same rule as the decision rule of \rt defenses (\eqref{eq:rt}).
It was also used by \citet{salman_provably_2019} where $\mathcal{L}$ is cross-entropy loss.
$H^{\mathrm{logits}}$ or an expected logits, is similar to $H^{\mathrm{softmax}}$ but without the softmax function to avoid potential vanishing gradients from softmax.

\begin{figure}[t!]
     \centering
     \begin{subfigure}[b]{0.23\textwidth}
         \centering
         \includegraphics[width=\textwidth]{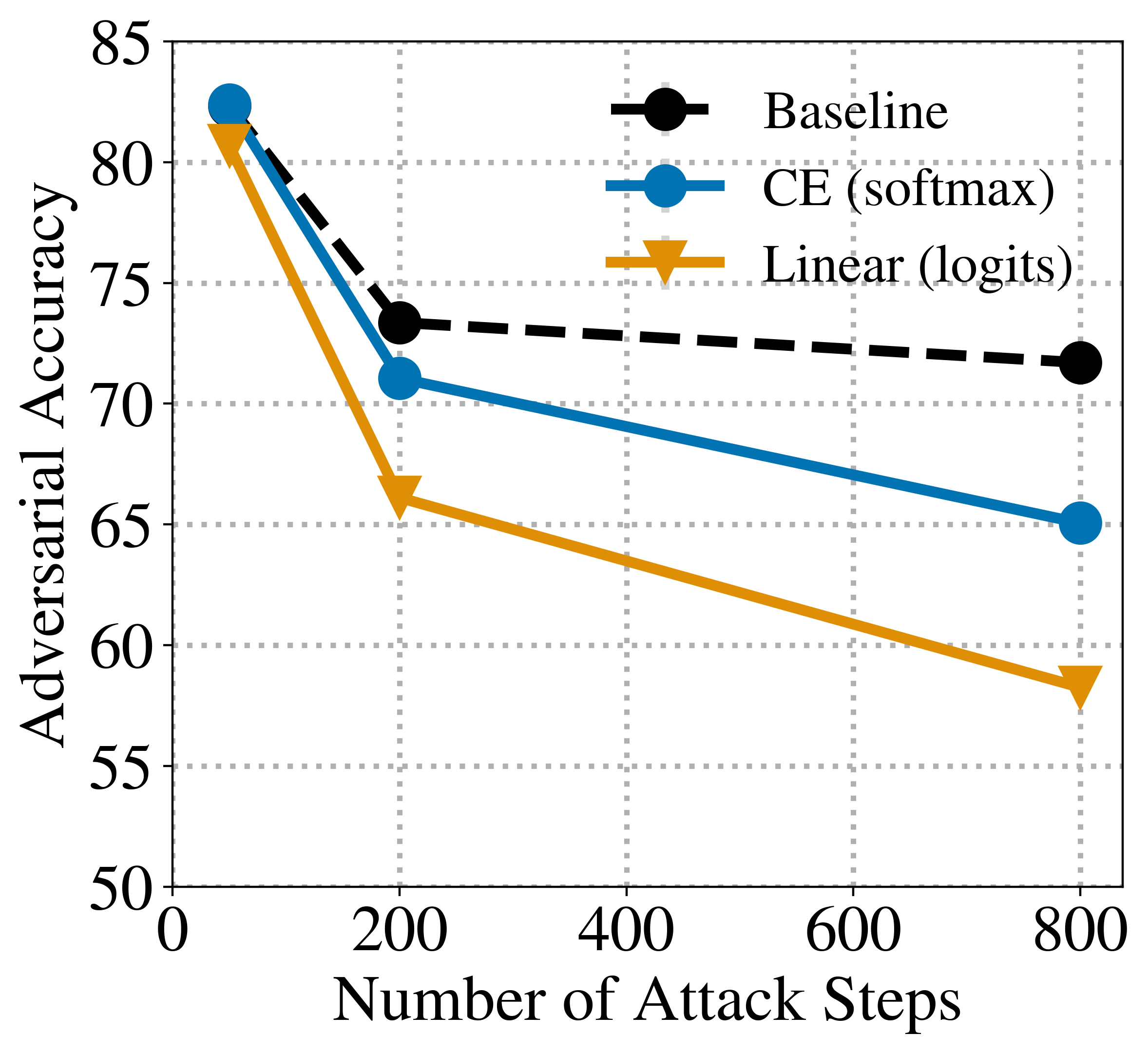}
         \includegraphics[width=\textwidth]{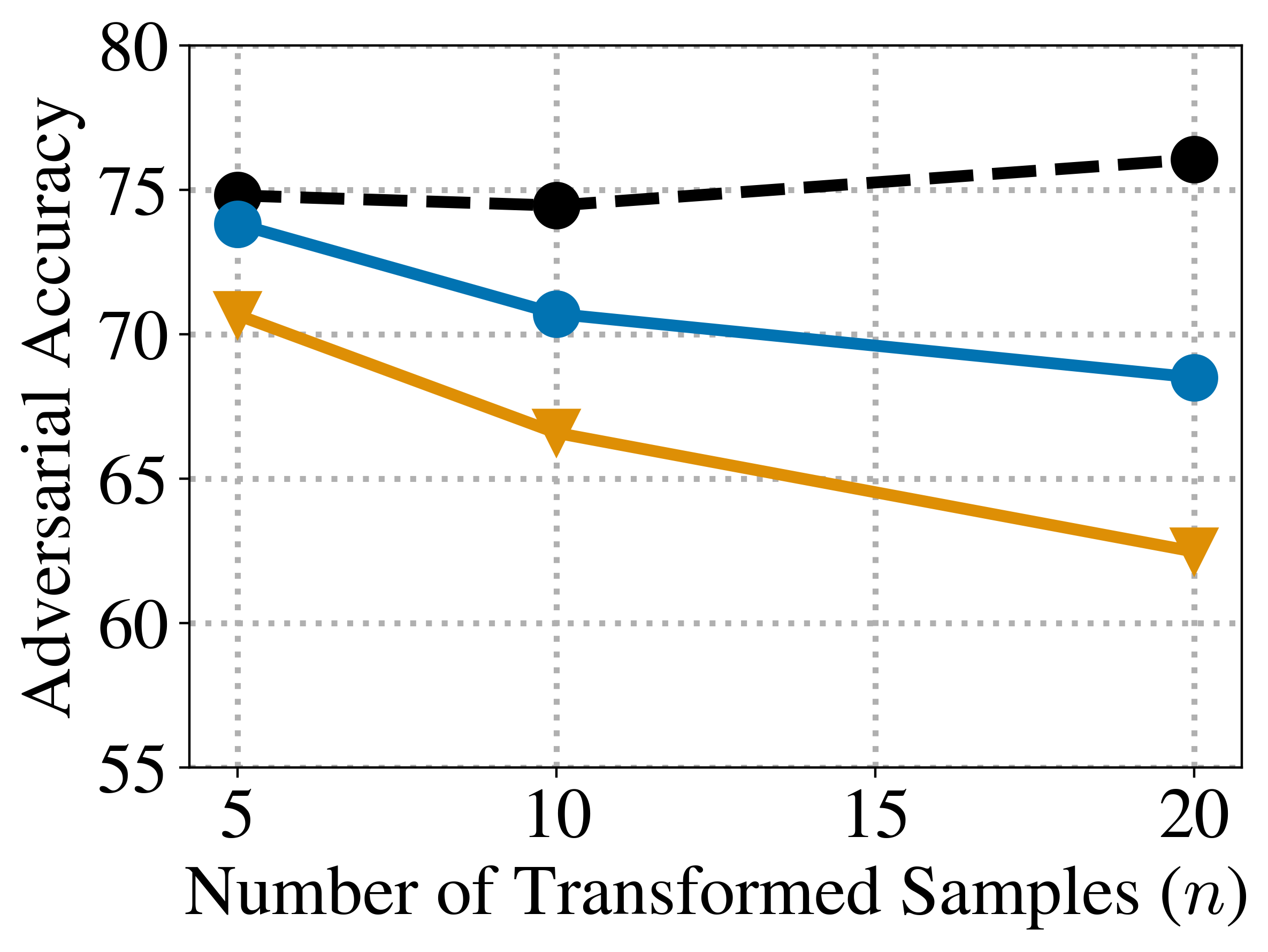}
         \caption{Comparison among loss functions and decision rules}
         \label{fig:img_attack_loss}
     \end{subfigure}
     \hfill
     \begin{subfigure}[b]{0.23\textwidth}
         \centering
         \includegraphics[width=\textwidth]{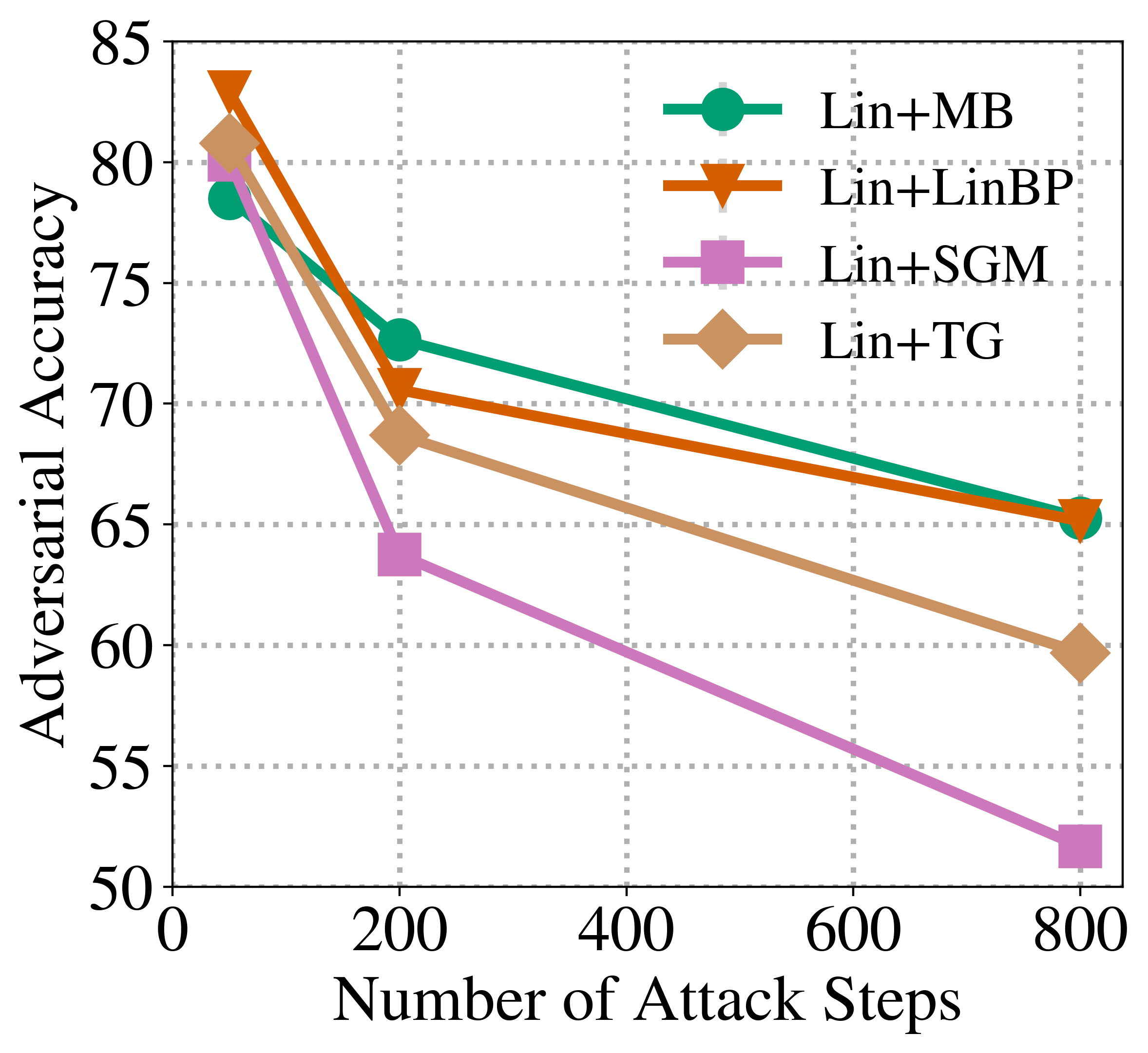}
         \includegraphics[width=\textwidth]{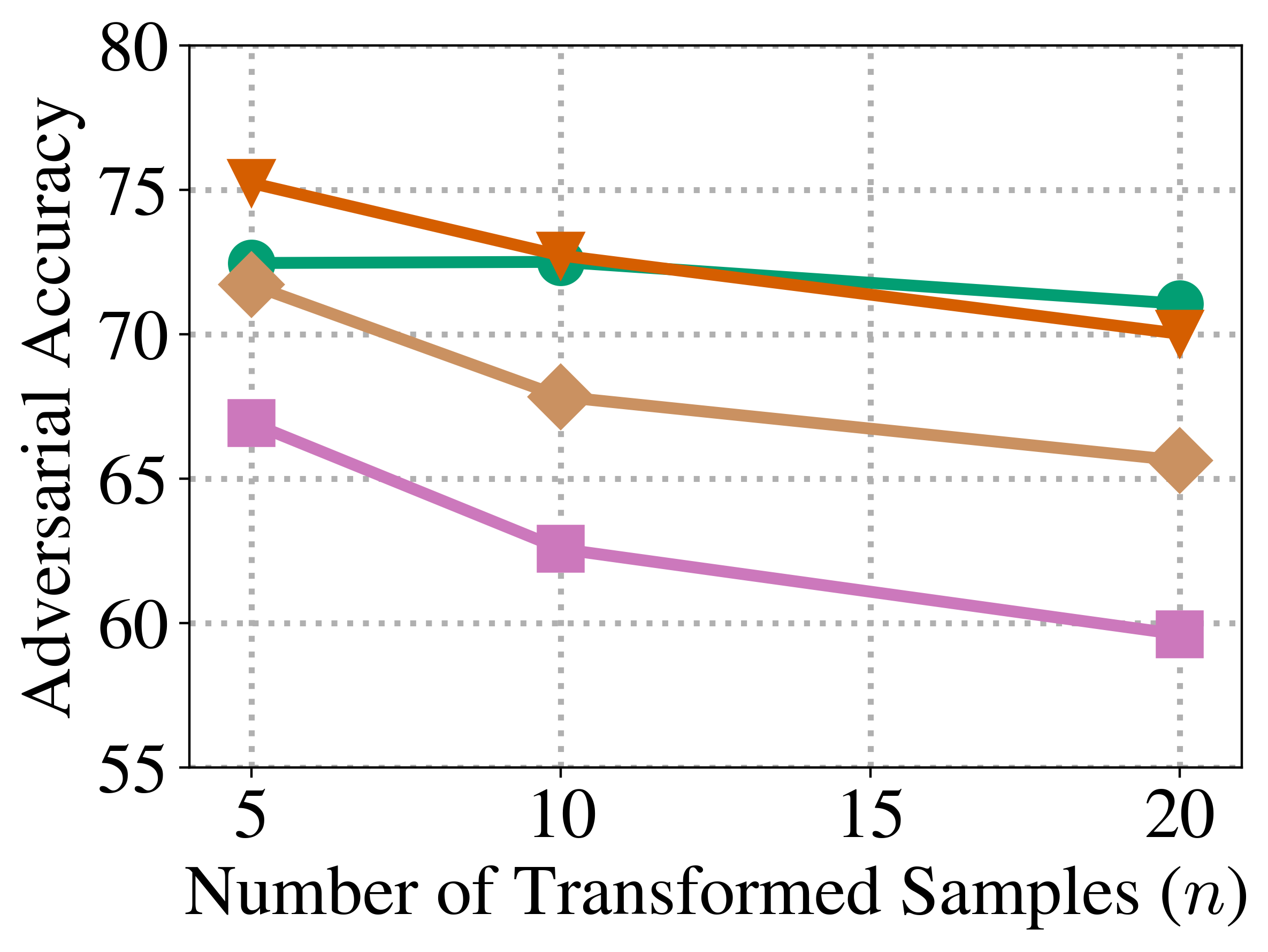}
         \caption{Comparison among transfer attack techniques}
         \label{fig:img_attack_ens}
     \end{subfigure}
    \caption{Comparison of PGD attack's effectiveness with (a) different loss functions and decision rules, and (b) different attack variants with improved transferability. The error bars are too small to see with the markers so we report the numerical results in Table~\ref{tab:main_attack}. 
    ``Baseline'' refers to EoT with CE loss in \eqref{eq:attack_eot}.
    }
    \label{fig:attack_loss_ens}
\end{figure}

In addition to the rules, we experiment with two choices of $\mathcal{L}$ commonly used for generating adversarial examples: cross-entropy loss (CE) and linear loss (Linear). 
The linear loss is defined as the difference between the largest logit of the wrong class and logit of the correct class:
\begin{align}
    \mathcal{L}_{\mathrm{Linear}}(x, y) &~\coloneqq~ \max_{j \ne y} F_j - F_y \\
    \text{where}~\;~ F &~=~ \Ex_{\theta \sim p(\theta)} \left[f\left(t(x; \theta) \right) \right]
\end{align}
The advantage of the linear loss is that its gradient estimates are unbiased, similarly to EoT, meaning that the expectation can be moved in front of $\mathcal{L}$ due to linearity.
However, this is not the case for CE loss.

\textbf{Attack evaluation and comparison.} 
We evaluate the attacks by their effectiveness in reducing the adversarial accuracy (lower means stronger attack) on the \rt defense obtained from Section~\ref{sec:bayesopt}.
In our setting, the adversarial examples are generated once and then used to compute the accuracy 10 times, each with a different random seed on the \rt defense. 
We report the average accuracy over these 10 runs together with the 95\%-confidence interval.
Alternatively, one can imagine a threat model that counts at least one misclassification among a certain number of trials as incorrect. 
This is an interesting and perhaps more realistic in some settings, but the optimal attack will be very different from EoT as we care a lot less about the expectation.
This, however, is outside of the scope of our work.

In \figref{fig:img_attack_loss}, we compare the effectiveness of four attacks, each using a different pair of losses and decision rules with varying numbers of PGD steps and samples $n$.
The widely used EoT method performs the worst of the four.
CE loss on mean softmax probability performs better than EoT, confirming the observation made by \citet{salman_provably_2019}.
Linear loss and CE loss on average logits are even better and are consistently the strongest attacks, across all hyperparameters.
For the rest of this paper, we adopt the linear loss with mean logits as the main objective function.

\begin{figure}
     \centering
     \includegraphics[width=0.37\textwidth]{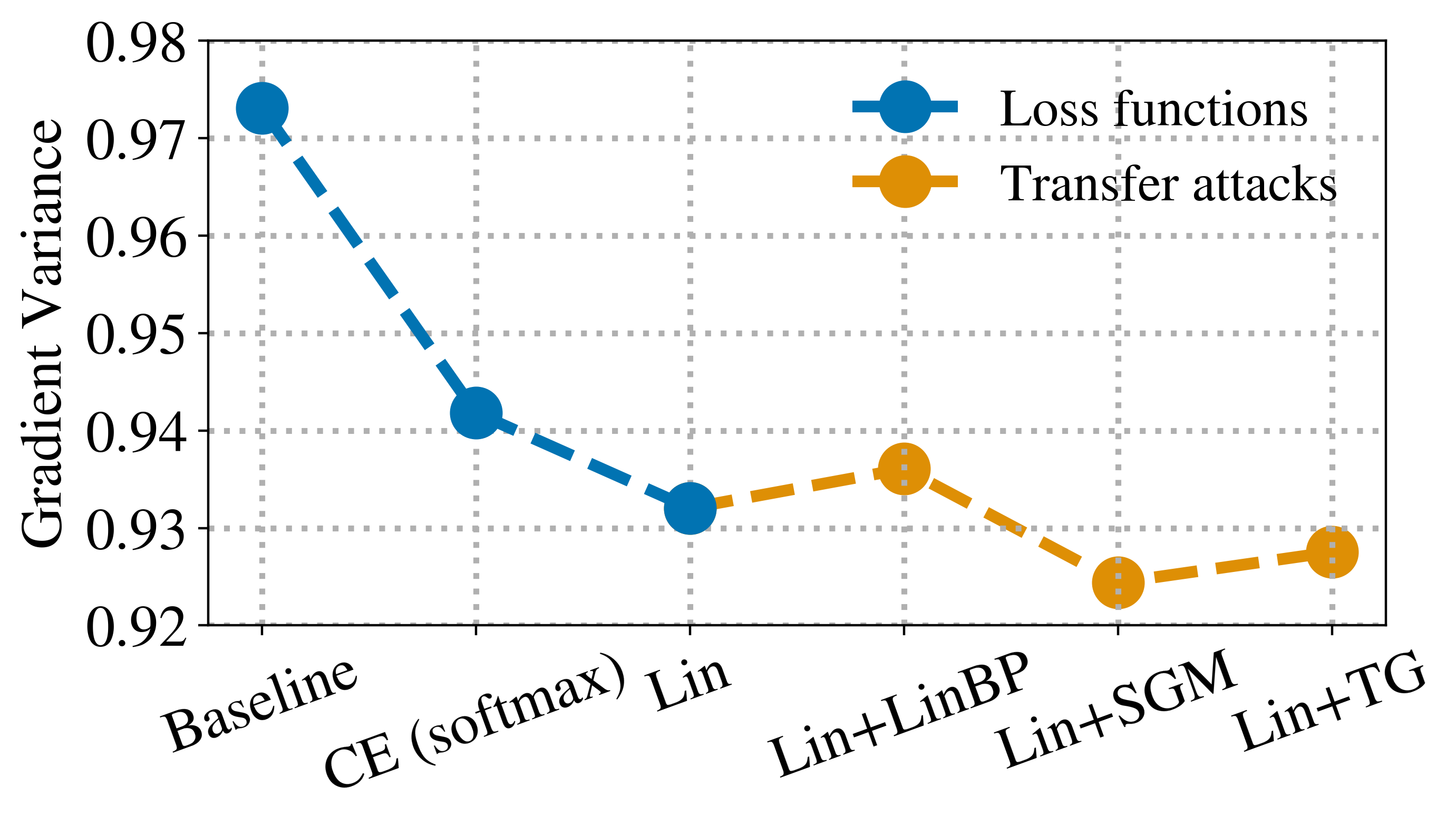}
     \caption{Comparison of dimension-normalized variance of the gradient estimates across (blue) different loss functions and decision rules and (yellow) transferability-improving attacks. Strong attacks are highly correlated with low variance of their gradient estimates, i.e., Lin+SGM. Note that Lin+MB or Momentum Boosting is not shown here because it does not modify the gradients.}
     \label{fig:main_var}
\end{figure}

\textbf{Connection to variance.}
As we predicted in Section~\ref{ssec:var_sgd}, a stronger attack directly corresponds to lower variance.
This hypothesis is confirmed by \figref{fig:main_var}.
For instance, the EoT baseline has the highest variance as well as the worst performance according to \figref{fig:atk_img_rand}.
On the other hand, the linear loss (Lin) has the lowest variance among the three loss functions (blue) and hence, it performs the best.
The other three points in orange will be covered in the next section.

\subsection{Ensemble and Transfer Attacks} \label{ssec:ensemble}

\rt defense can be regarded as an ensemble of neural networks with each member sharing the same parameters but applying different sets of transformations to the input (i.e., different $\theta$'s from random sampling).
Consequently, we may view a white-box attack on \rt defenses as a ``partial'' black-box attack on an ensemble of (infinitely) many models where the adversary wishes to ``transfer'' adversarial examples generated on some subset of the members to another unseen subset.

Given this interpretation, we apply four techniques designed to enhance the transferability of adversarial examples to improve the attack success rate on \rt defense.
The techniques include momentum boosting (MB)~\cite{dong_boosting_2018}, modifying backward passes by ignoring non-linear activation (LinBP)~\cite{guo_backpropagating_2020} or by emphasizing the gradient through skip connections of ResNets more than through the residual block (SGM)~\cite{wu_skip_2020}, and simply using a targeted attack with the linear loss function (TG)~\cite{zhao_success_2021}.
In \figref{fig:img_attack_ens}, we compare these techniques combined with the best performing loss and decision rule from Section~\ref{ssec:adv_obj} (i.e., the linear loss on logits).
Only SGM improves the attack success rate at all settings while the rest result in weaker attacks than the one without any of the techniques (denoted by ``Linear (logits)'' in \figref{fig:img_attack_loss}).

SGM essentially normalizes the gradients and scales ones from the residual blocks by some constant less than 1 (we use $0.5$) to reduce its influence and prioritize the gradients from the skip connection. 
\citet{wu_skip_2020} explain that SGM leads to better transferability because gradients through skip connections preserve ``low-level information'' which tends to transfer better.
Intuitively, this agrees with our variance explanation as
the increased transferability implies a stronger agreement among gradient samples and hence, less spread or lower variance.

\subsection{Stochastic Optimization Algorithm} \label{ssec:optimizer}

While most attacks on deterministic models can use naive PGD to solve \eqref{eq:adv} effectively, this is not the case for stochastic models like the \rt defense. 
Here, the adversary only has access to noisy estimates of the gradients, making it a strictly more difficult problem, and techniques used in the deterministic case may no longer apply.

\begin{figure}[t!]
    \centering
    \includegraphics[width=0.36\textwidth]{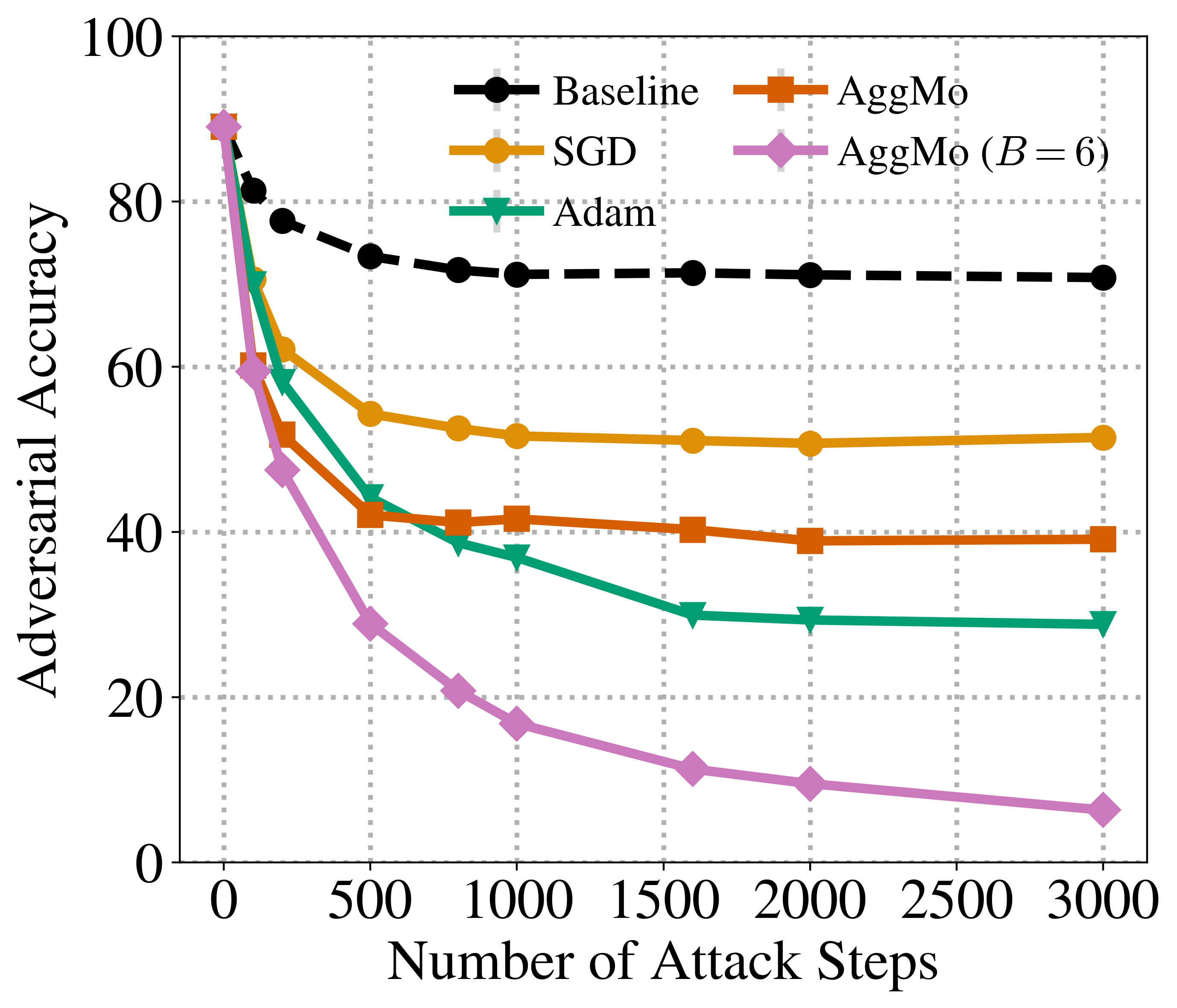}~
    \caption{Comparison of the optimizers for attacking an \rt defense with $\epsilon=16/255, n=10$ on Imagenette dataset. All but the baseline (CE loss with EoT) use the linear loss with SGM, and all but AggMo~($B=6$) use the default hyperparameters. AggMo with $B=6$ outperforms the other algorithms in terms of both the convergence rate and the final adversarial accuracy obtained. This result is not very sensitive to $B$ as any sufficiently large value ($\ge 4$) yields the same outcome.}
    \label{fig:atk_img_rand}
\end{figure}

As mentioned in Section~\ref{ssec:var_sgd}, high-variance gradient estimates undermine the convergence rate of SGD.
Thus, the attack should benefit from optimization techniques aimed at reducing the variance or speeding up the convergence of SGD.
We first experiment with common optimizers such as SGD and Adam~\citep{kingma_adam_2015} with different hyperparameters, e.g., momentum, Nesterov acceleration, and learning rate schedules, to find the best setting for the linear loss with SGM.
Based on this experiment, we found that a momentum term with an appropriate damping constant plays an important role in the attack success rate.
Momentum is also well-known to accelerate and stabilize training of neural networks~\citep{sutskever_importance_2013a}.
\figref{fig:atk_img_rand_sgd} reports adversarial accuracy at varying attack iterations and indicates that higher momentum constant leads to faster convergence and a higher attack success rate.
However, the results seem highly sensitive to this momentum constant which also varies from one setting to another (e.g., number or types of transformations, dataset, etc.).

To mitigate this issue, we introduce another optimizer. AggMo is exactly designed to be less sensitive to choices of the damping coefficient by aggregating $B$ momentum terms with different constants instead of one~\citep{lucas_aggregated_2019}.
After only a few tries, we found a wide range of values of $B$ where AggMo outperforms SGD with a fine-tuned momentum constant (see \figref{fig:atk_img_rand_aggmo}).
\figref{fig:atk_img_rand} compares the attacks using different choices of the optimizers to the baseline EoT attack.
Here, the baseline can only reduce the adversarial accuracy from $89\%$ to $70\%$ while \textbf{our best attack manages to reach $\bm{6\%}$ or over $\bm{4.3\times}$ improvement.}
This concludes that the optimizer plays a crucial role in the success of the attack, and \textbf{the \rt defense, even with a carefully and systematically chosen transformation hyperparameters, is not robust against adversarial examples.}

Furthermore, we note that without our loss function and only using AggMo, the accuracy only goes down to $23\%$ at a much slower rate.
Conversely, when the linear loss and SGM are used with SGD (no momentum), the accuracy drops to $51\%$. 
This signifies that all three techniques we deploy play important roles to the attack's effectiveness.

\subsection{Comparison with AutoAttack}

AutoAttack~\citep{croce_reliable_2020} was proposed as a standardized benchmark for evaluating deterministic defenses against adversarial examples.
It uses an ensemble of four different attacks that cover weaknesses of one another, one of which does not use gradients.
AutoAttack has been proven to be one of the strongest attack currently and is capable of catching defenses with false robustness caused by gradient obfuscation~\citep{athalye_obfuscated_2018}.

While not particularly designed for stochastic models, AutoAttack can be used to evaluate them when combined with EoT.
We report the accuracy on adversarial examples generated on AutoAttack with all default hyperparameters in the ``standard'' mode and 10-sample EoT in Table~\ref{tab:attack_compare}.
AutoAttack performs worse than the baseline EoT and our attack on both Imagenette and CIFAR-10 by a large margin.
One of the reasons is that AutoAttack is optimized for efficiency and so each of its attacks is usually terminated once a misclassification occurs.
This is applicable to deterministic models, but for stochastic ones such as an \rt defense, the adversary is better off finding the adversarial examples that maximize the expected loss instead of ones that are misclassified once.

To take this property into account, we include the accuracy reported by AutoAttack that treats a sample as incorrect if it is misclassified at least \emph{once} throughout the entire process.
For Imagenette, the accuracies after each of the four attacks (APGD-CE, APGD-T, FAB, and Square) is applied sequentially are $82.03$, $78.81$, $78.03$, and $77.34$, respectively.
Note that this is a one-time evaluation so there is no error bar here.
Needless to say, the adversarial accuracy computed this way is strictly lower than the one we reported in Table~\ref{tab:attack_compare} and violates our threat model. 
However, it is still higher than that of the baseline EoT and our attack, suggesting that AutoAttack is ineffective against randomized models like \rt defenses.
AutoAttack also comes with a ``random'' mode for randomized models which only use APGD-CE and APGD-DLR with 20-sample EoT.
The adversarial accuracies obtained from this mode are $85.62$ and $83.83$ or
$88.62 \pm 0.46$ for single-pass evaluation as in Table~\ref{tab:attack_compare}. This random mode performs worse than the standard version.

\section{Combining with Adversarial Training} \label{sec:combine_at}

\begin{table*}[t!]
\small
\centering
\caption{Comparison of \rt and \art defenses to prior robust deterministic models and a normally trained model. Both the \rt and the \art models on Imagenette lack the adversarial robustness. Conversely, the \rt defense on CIFAR-10 does bring substantial robustness, and combining it with adversarial training boosts the adversarial accuracy further. Nonetheless, they still fall behind the previously proposed deterministic models including \citet{madry_deep_2018} and \citet{zhang_theoretically_2019}. The largest number in each column is in bold.}
\label{tab:adv_compare}
\begin{tabular}{@{}lrrrr@{}}
\toprule
\multirow{2}{*}{Defenses} & \multicolumn{2}{c}{Imagenette} & \multicolumn{2}{c}{CIFAR-10} \\
\cmidrule(lr){2-3} \cmidrule(lr){4-5}
& Clean Accuracy & Adv. Accuracy & Clean Accuracy & Adv. Accuracy \\ \midrule
Normal model       & $\bm{95.41}$    & $0.00$ & $\bm{95.10}$ & $0.00$ \\
\citet{madry_deep_2018} & $78.25$    & $\bm{37.10}$  & $81.90$ & $45.30$ \\
\citet{zhang_theoretically_2019}  & $87.43$    & $33.19$ & $81.26$ & $\bm{46.89}$  \\
\rt defense & $89.04 \pm 0.34$ & $6.34 \pm 0.35$ & $81.12 \pm 0.54$ & $29.91 \pm 0.35$ \\
\art defense & $88.83 \pm 0.26$ & $8.68\pm 0.52$ & $80.69 \pm 0.66$ & $41.30 \pm 0.49$ \\
\bottomrule
\end{tabular}
\end{table*}

To deepen our investigation, we explore the possibility of combining \rt defense with adversarial training.
However, this is a challenging problem on its own.
For normal deterministic models, 10-step PGD is sufficient for reaching adversarial accuracy close to best known attack or the optimal adversarial accuracy.
However, this is not the case for \rt defenses as even our new attack still requires more than one thousand iterations before the adversarial accuracy starts to plateau.
Ultimately, the robustness of adversarially trained models largely depends on the strength of the attack used to generate the adversarial examples, and using a weak attack means that the obtained model will not be robust.
A similar phenomenon is observed by \citet{tramer_ensemble_2018} and \citet{wong_fast_2020} where an adversarially trained model overfits to the weak FGSM attacks but has shown to be non-robust with the accurate evaluation. 
To test this hypothesis, we adversarially train the \rt defense from Section~\ref{sec:bayesopt} using our new attack with 50 iterations (already $5\times$ the common number of steps) and call this defense ''\art.''
The attack step size is also adjusted accordingly to $\epsilon / 8$.

In Table~\ref{tab:adv_compare}, we confirm that training \art this way results in a model with virtually no robustness improvement over the normal \rt on Imagenette.
On the other hand, the \art trained on CIFAR-10 proves to be more promising even though it is still not as robust as deterministic models trained with adversarial training or TRADES~\citep{zhang_theoretically_2019}.
Based on this result, \textbf{we conclude that a stronger attack on \rt defenses that converge within a much fewer iterations will be necessary to make adversarial training successful.}
In theory, it might be possible to achieve a robust \rt model with 1,000-step attack on Imagenette, but this is too computationally intensive for us to verify, and it will not to scale to any realistic setting.

\section{Conclusion}

While recent papers report state-of-the-art robustness with \rt defenses, our evaluations show that \rt generally under-performs existing defenses like adversarial training when met with a stronger attack, even after fine-tuning the hyperparameters of the defense. 
Through our experiments, we found that non-differentiability and high-variance gradients can seriously inhibit adversarial optimization, so we recommend using only differentiable transformations along with their exact gradients in the evaluation of future \rt defenses.
In this setting, we propose a new state-of-the-art attack that improves significantly over the baseline (PGD with EoT) and show that \rt defenses as well as their adversarially trained counterparts are not as robust to adversarial examples as they were previously believed to be.

\section*{Acknowledgements}
We would like to thank Jonathan Shewchuk for the feedback on the paper.
This research was supported by the Hewlett Foundation through the Center for Long-Term Cybersecurity (CLTC), by the Berkeley Deep Drive project, by the National Science Foundation under Award CCF-1909204, and by generous gifts from Open Philanthropy and Google Cloud Research Credits program under Award GCP19980904.

\bibliographystyle{icml2022}
\bibliography{bib/additional.bib,bib/reference.bib}

\begin{thebibliography}{48}
\providecommand{\natexlab}[1]{#1}
\providecommand{\url}[1]{\texttt{#1}}
\expandafter\ifx\csname urlstyle\endcsname\relax
  \providecommand{\doi}[1]{doi: #1}\else
  \providecommand{\doi}{doi: \begingroup \urlstyle{rm}\Url}\fi

\bibitem[Athalye et~al.(2018{\natexlab{a}})Athalye, Carlini, and
  Wagner]{athalye_obfuscated_2018}
Athalye, A., Carlini, N., and Wagner, D.
\newblock Obfuscated gradients give a false sense of security: Circumventing
  defenses to adversarial examples.
\newblock In Dy, J. and Krause, A. (eds.), \emph{Proceedings of the 35th
  International Conference on Machine Learning}, volume~80 of \emph{Proceedings
  of Machine Learning Research}, pp.\  274--283, {Stockholmsm\"assan, Stockholm
  Sweden}, July 2018{\natexlab{a}}. {PMLR}.

\bibitem[Athalye et~al.(2018{\natexlab{b}})Athalye, Engstrom, Ilyas, and
  Kwok]{athalye_synthesizing_2018}
Athalye, A., Engstrom, L., Ilyas, A., and Kwok, K.
\newblock Synthesizing robust adversarial examples.
\newblock In Dy, J. and Krause, A. (eds.), \emph{Proceedings of the 35th
  International Conference on Machine Learning}, volume~80 of \emph{Proceedings
  of Machine Learning Research}, pp.\  284--293, {Stockholmsm\"assan, Stockholm
  Sweden}, July 2018{\natexlab{b}}. {PMLR}.

\bibitem[Bender et~al.(2020)Bender, Li, Shi, Reiter, and
  Oliva]{bender_defense_2020}
Bender, C., Li, Y., Shi, Y., Reiter, M.~K., and Oliva, J.
\newblock Defense through diverse directions.
\newblock In III, H.~D. and Singh, A. (eds.), \emph{Proceedings of the 37th
  International Conference on Machine Learning}, volume 119 of
  \emph{Proceedings of Machine Learning Research}, pp.\  756--766. {PMLR}, July
  2020.

\bibitem[Bergstra et~al.(2013)Bergstra, Yamins, and Cox]{bergstra_making_2013}
Bergstra, J., Yamins, D., and Cox, D.~D.
\newblock Making a science of model search: Hyperparameter optimization in
  hundreds of dimensions for vision architectures.
\newblock In \emph{Proceedings of the 30th International Conference on
  International Conference on Machine Learning - Volume 28}, {{ICML}}'13, pp.\
  I--115--I--123, {Atlanta, GA, USA}, 2013. {JMLR.org}.

\bibitem[Bernstein et~al.(2018)Bernstein, Wang, Azizzadenesheli, and
  Anandkumar]{bernstein_signsgd_2018}
Bernstein, J., Wang, Y.-X., Azizzadenesheli, K., and Anandkumar, A.
\newblock {{signSGD}}: Compressed optimisation for non-convex problems.
\newblock In Dy, J. and Krause, A. (eds.), \emph{Proceedings of the 35th
  International Conference on Machine Learning}, volume~80 of \emph{Proceedings
  of Machine Learning Research}, pp.\  560--569. {PMLR}, July 2018.

\bibitem[Biggio et~al.(2013)Biggio, Corona, Maiorca, Nelson, {\v S}rndi{\'c},
  Laskov, Giacinto, and Roli]{biggio_evasion_2013}
Biggio, B., Corona, I., Maiorca, D., Nelson, B., {\v S}rndi{\'c}, N., Laskov,
  P., Giacinto, G., and Roli, F.
\newblock Evasion attacks against machine learning at test time.
\newblock In Blockeel, H., Kersting, K., Nijssen, S., and {\v Z}elezn{\'y}, F.
  (eds.), \emph{Machine {{Learning}} and {{Knowledge Discovery}} in
  {{Databases}}}, pp.\  387--402, {Berlin, Heidelberg}, 2013. {Springer Berlin
  Heidelberg}.
\newblock ISBN 978-3-642-40994-3.

\bibitem[Brochu et~al.(2010)Brochu, Cora, and {de
  Freitas}]{brochu_tutorial_2010}
Brochu, E., Cora, V.~M., and {de Freitas}, N.
\newblock A tutorial on bayesian optimization of expensive cost functions, with
  application to active user modeling and hierarchical reinforcement learning.
\newblock \emph{arXiv:1012.2599 [cs]}, December 2010.

\bibitem[Chen et~al.(2021)Chen, Wu, Guo, Liang, and Jha]{chen_evaluating_2021}
Chen, J., Wu, X., Guo, Y., Liang, Y., and Jha, S.
\newblock Towards evaluating the robustness of neural networks learned by
  transduction.
\newblock \emph{arXiv:2110.14735 [cs]}, October 2021.

\bibitem[Cohen et~al.(2019)Cohen, Rosenfeld, and Kolter]{cohen_certified_2019}
Cohen, J., Rosenfeld, E., and Kolter, Z.
\newblock Certified adversarial robustness via randomized smoothing.
\newblock In Chaudhuri, K. and Salakhutdinov, R. (eds.), \emph{Proceedings of
  the 36th International Conference on Machine Learning}, volume~97 of
  \emph{Proceedings of Machine Learning Research}, pp.\  1310--1320. {PMLR},
  June 2019.

\bibitem[Croce \& Hein(2020)Croce and Hein]{croce_reliable_2020}
Croce, F. and Hein, M.
\newblock Reliable evaluation of adversarial robustness with an ensemble of
  diverse parameter-free attacks.
\newblock In III, H.~D. and Singh, A. (eds.), \emph{Proceedings of the 37th
  International Conference on Machine Learning}, volume 119 of
  \emph{Proceedings of Machine Learning Research}, pp.\  2206--2216. {PMLR},
  July 2020.

\bibitem[Dhillon et~al.(2018)Dhillon, Azizzadenesheli, Bernstein, Kossaifi,
  Khanna, Lipton, and Anandkumar]{dhillon_stochastic_2018}
Dhillon, G.~S., Azizzadenesheli, K., Bernstein, J.~D., Kossaifi, J., Khanna,
  A., Lipton, Z.~C., and Anandkumar, A.
\newblock Stochastic activation pruning for robust adversarial defense.
\newblock In \emph{International Conference on Learning Representations}, 2018.

\bibitem[Dong et~al.(2018)Dong, Liao, Pang, Su, Zhu, Hu, and
  Li]{dong_boosting_2018}
Dong, Y., Liao, F., Pang, T., Su, H., Zhu, J., Hu, X., and Li, J.
\newblock Boosting adversarial attacks with momentum.
\newblock In \emph{2018 {{IEEE}}/{{CVF Conference}} on {{Computer Vision}} and
  {{Pattern Recognition}} ({{CVPR}})}. {IEEE}, March 2018.

\bibitem[Falkner et~al.(2018)Falkner, Klein, and Hutter]{falkner_bohb_2018}
Falkner, S., Klein, A., and Hutter, F.
\newblock {{BOHB}}: Robust and efficient hyperparameter optimization at scale.
\newblock In Dy, J. and Krause, A. (eds.), \emph{Proceedings of the 35th
  International Conference on Machine Learning}, volume~80 of \emph{Proceedings
  of Machine Learning Research}, pp.\  1437--1446, {Stockholmsm\"assan,
  Stockholm Sweden}, July 2018. {PMLR}.

\bibitem[Ghadimi \& Lan(2013)Ghadimi and Lan]{ghadimi_stochastic_2013}
Ghadimi, S. and Lan, G.
\newblock Stochastic first- and zeroth-order methods for nonconvex stochastic
  programming.
\newblock \emph{SIAM Journal on Optimization}, 23\penalty0 (4):\penalty0
  2341--2368, January 2013.
\newblock ISSN 1052-6234.
\newblock \doi{10.1137/120880811}.

\bibitem[Goodfellow et~al.(2015)Goodfellow, Shlens, and
  Szegedy]{goodfellow_explaining_2015}
Goodfellow, I., Shlens, J., and Szegedy, C.
\newblock Explaining and harnessing adversarial examples.
\newblock In \emph{International Conference on Learning Representations}, 2015.

\bibitem[Guo et~al.(2018)Guo, Rana, Cisse, and {van der
  Maaten}]{guo_countering_2018}
Guo, C., Rana, M., Cisse, M., and {van der Maaten}, L.
\newblock Countering adversarial images using input transformations.
\newblock In \emph{International Conference on Learning Representations}, 2018.

\bibitem[Guo et~al.(2020)Guo, Li, and Chen]{guo_backpropagating_2020}
Guo, Y., Li, Q., and Chen, H.
\newblock Backpropagating linearly improves transferability of adversarial
  examples.
\newblock In \emph{{{NeurIPS}}}, 2020.

\bibitem[Gupta et~al.(2020)Gupta, Dube, and Verma]{gupta_improving_2020}
Gupta, S., Dube, P., and Verma, A.
\newblock Improving the affordability of robustness training for {{DNNs}}.
\newblock In \emph{Proceedings of the {{IEEE}}/{{CVF}} Conference on Computer
  Vision and Pattern Recognition ({{CVPR}}) Workshops}, June 2020.

\bibitem[He et~al.(2016{\natexlab{a}})He, Zhang, Ren, and Sun]{he_deep_2016}
He, K., Zhang, X., Ren, S., and Sun, J.
\newblock Deep residual learning for image recognition.
\newblock In \emph{2016 {{IEEE}} Conference on Computer Vision and Pattern
  Recognition ({{CVPR}})}, pp.\  770--778, 2016{\natexlab{a}}.
\newblock \doi{10.1109/CVPR.2016.90}.

\bibitem[He et~al.(2016{\natexlab{b}})He, Zhang, Ren, and
  Sun]{he_identity_2016}
He, K., Zhang, X., Ren, S., and Sun, J.
\newblock Identity mappings in deep residual networks.
\newblock In \emph{European Conference on Computer Vision}, pp.\  630--645.
  {Springer}, 2016{\natexlab{b}}.

\bibitem[He et~al.(2018)He, Li, and Song]{he_decision_2018}
He, W., Li, B., and Song, D.
\newblock Decision boundary analysis of adversarial examples.
\newblock In \emph{International Conference on Learning Representations}, 2018.

\bibitem[He et~al.(2019)He, Rakin, and Fan]{he_parametric_2019}
He, Z., Rakin, A.~S., and Fan, D.
\newblock Parametric noise injection: {{Trainable}} randomness to improve deep
  neural network robustness against adversarial attack.
\newblock In \emph{Proceedings of the {{IEEE}} Conference on Computer Vision
  and Pattern Recognition}, pp.\  588--597, 2019.

\bibitem[Howard(2021)]{howard_fastai_2021}
Howard, J.
\newblock Fastai/imagenette.
\newblock fast.ai, March 2021.

\bibitem[Kandasamy et~al.(2020)Kandasamy, Vysyaraju, Neiswanger, Paria,
  Collins, Schneider, Poczos, and Xing]{kandasamy_tuning_2020}
Kandasamy, K., Vysyaraju, K.~R., Neiswanger, W., Paria, B., Collins, C.~R.,
  Schneider, J., Poczos, B., and Xing, E.~P.
\newblock Tuning hyperparameters without grad students: Scalable and robust
  bayesian optimisation with dragonfly.
\newblock \emph{Journal of Machine Learning Research}, 21\penalty0
  (81):\penalty0 1--27, 2020.

\bibitem[Kingma \& Ba(2015)Kingma and Ba]{kingma_adam_2015}
Kingma, D.~P. and Ba, J.
\newblock Adam: {{A}} method for stochastic optimization.
\newblock In Bengio, Y. and LeCun, Y. (eds.), \emph{3rd International
  Conference on Learning Representations, {{ICLR}} 2015, San Diego, {{CA}},
  {{USA}}, May 7-9, 2015, Conference Track Proceedings}, 2015.

\bibitem[Lecuyer et~al.(2019)Lecuyer, Atlidakis, Geambasu, Hsu, and
  Jana]{lecuyer_certified_2019}
Lecuyer, M., Atlidakis, V., Geambasu, R., Hsu, D., and Jana, S.
\newblock Certified robustness to adversarial examples with differential
  privacy.
\newblock In \emph{2019 {{IEEE}} Symposium on Security and Privacy ({{SP}})},
  pp.\  656--672, 2019.
\newblock \doi{10.1109/SP.2019.00044}.

\bibitem[Liaw et~al.(2018)Liaw, Liang, Nishihara, Moritz, Gonzalez, and
  Stoica]{liaw_tune_2018}
Liaw, R., Liang, E., Nishihara, R., Moritz, P., Gonzalez, J.~E., and Stoica, I.
\newblock Tune: A research platform for distributed model selection and
  training.
\newblock \emph{arXiv preprint arXiv:1807.05118}, 2018.

\bibitem[Liu et~al.(2018)Liu, Cheng, Zhang, and Hsieh]{liu_robust_2018}
Liu, X., Cheng, M., Zhang, H., and Hsieh, C.-J.
\newblock Towards robust neural networks via random self-ensemble.
\newblock In \emph{{{ECCV}} (7)}, pp.\  381--397, 2018.

\bibitem[Liu et~al.(2019)Liu, Li, Wu, and Hsieh]{liu_advbnn_2019}
Liu, X., Li, Y., Wu, C., and Hsieh, C.-J.
\newblock Adv-{{BNN}}: {{Improved}} adversarial defense through robust bayesian
  neural network.
\newblock In \emph{International Conference on Learning Representations}, 2019.

\bibitem[Loshchilov \& Hutter(2017)Loshchilov and Hutter]{loshchilov_sgdr_2017}
Loshchilov, I. and Hutter, F.
\newblock {{SGDR}}: Stochastic gradient descent with warm restarts.
\newblock In \emph{International Conference on Learning Representations}, 2017.

\bibitem[Lucas et~al.(2019)Lucas, Sun, Zemel, and
  Grosse]{lucas_aggregated_2019}
Lucas, J., Sun, S., Zemel, R., and Grosse, R.
\newblock Aggregated momentum: Stability through passive damping.
\newblock In \emph{International Conference on Learning Representations}, 2019.

\bibitem[Madry et~al.(2018)Madry, Makelov, Schmidt, Tsipras, and
  Vladu]{madry_deep_2018}
Madry, A., Makelov, A., Schmidt, L., Tsipras, D., and Vladu, A.
\newblock Towards deep learning models resistant to adversarial attacks.
\newblock In \emph{International Conference on Learning Representations}, 2018.

\bibitem[Nogueira(2014)]{nogueira_bayesian_2014}
Nogueira, F.
\newblock Bayesian {{Optimization}}: {{Open}} source constrained global
  optimization tool for {{Python}}.
\newblock 2014.

\bibitem[Raff et~al.(2019)Raff, Sylvester, Forsyth, and
  McLean]{raff_barrage_2019}
Raff, E., Sylvester, J., Forsyth, S., and McLean, M.
\newblock Barrage of random transforms for adversarially robust defense.
\newblock In \emph{2019 {{IEEE}}/{{CVF Conference}} on {{Computer Vision}} and
  {{Pattern Recognition}} ({{CVPR}})}, pp.\  6521--6530, {Long Beach, CA, USA},
  June 2019. {IEEE}.
\newblock ISBN 978-1-72813-293-8.
\newblock \doi{10.1109/CVPR.2019.00669}.

\bibitem[Salman et~al.(2019)Salman, Li, Razenshteyn, Zhang, Zhang, Bubeck, and
  Yang]{salman_provably_2019}
Salman, H., Li, J., Razenshteyn, I., Zhang, P., Zhang, H., Bubeck, S., and
  Yang, G.
\newblock Provably robust deep learning via adversarially trained smoothed
  classifiers.
\newblock In Wallach, H., Larochelle, H., Beygelzimer, A., {dAlch{\'e}-Buc},
  F., Fox, E., and Garnett, R. (eds.), \emph{Advances in Neural Information
  Processing Systems}, volume~32. {Curran Associates, Inc.}, 2019.

\bibitem[Shin \& Song(2017)Shin and Song]{shin_jpegresistant_2017}
Shin, R. and Song, D.
\newblock {{JPEG-resistant}} adversarial images.
\newblock In \emph{Machine {{Learning}} and {{Computer Security Workshop}}
  (Co-Located with {{NeurIPS}} 2017)}, {Long Beach, CA, USA}, 2017.

\bibitem[Snoek et~al.(2012)Snoek, Larochelle, and Adams]{snoek_practical_2012}
Snoek, J., Larochelle, H., and Adams, R.~P.
\newblock Practical bayesian optimization of machine learning algorithms.
\newblock In Pereira, F., Burges, C. J.~C., Bottou, L., and Weinberger, K.~Q.
  (eds.), \emph{Advances in Neural Information Processing Systems}, volume~25.
  {Curran Associates, Inc.}, 2012.

\bibitem[Sutskever et~al.(2013)Sutskever, Martens, Dahl, and
  Hinton]{sutskever_importance_2013a}
Sutskever, I., Martens, J., Dahl, G., and Hinton, G.
\newblock On the importance of initialization and momentum in deep learning.
\newblock In Dasgupta, S. and McAllester, D. (eds.), \emph{Proceedings of the
  30th International Conference on Machine Learning}, volume~28 of
  \emph{Proceedings of Machine Learning Research}, pp.\  1139--1147, {Atlanta,
  Georgia, USA}, June 2013. {PMLR}.

\bibitem[Szegedy et~al.(2014)Szegedy, Zaremba, Sutskever, Bruna, Erhan,
  Goodfellow, and Fergus]{szegedy_intriguing_2014}
Szegedy, C., Zaremba, W., Sutskever, I., Bruna, J., Erhan, D., Goodfellow, I.,
  and Fergus, R.
\newblock Intriguing properties of neural networks.
\newblock In \emph{International Conference on Learning Representations}, 2014.

\bibitem[Tram{\`e}r et~al.(2018)Tram{\`e}r, Kurakin, Papernot, Goodfellow,
  Boneh, and McDaniel]{tramer_ensemble_2018}
Tram{\`e}r, F., Kurakin, A., Papernot, N., Goodfellow, I., Boneh, D., and
  McDaniel, P.
\newblock Ensemble adversarial training: Attacks and defenses.
\newblock In \emph{International Conference on Learning Representations}, 2018.

\bibitem[Tramer et~al.(2020)Tramer, Carlini, Brendel, and
  Madry]{tramer_adaptive_2020}
Tramer, F., Carlini, N., Brendel, W., and Madry, A.
\newblock On adaptive attacks to adversarial example defenses.
\newblock In Larochelle, H., Ranzato, M., Hadsell, R., Balcan, M.~F., and Lin,
  H. (eds.), \emph{Advances in Neural Information Processing Systems},
  volume~33, pp.\  1633--1645. {Curran Associates, Inc.}, 2020.

\bibitem[Tsipras et~al.(2019)Tsipras, Santurkar, Engstrom, Turner, and
  Madry]{tsipras_robustness_2019}
Tsipras, D., Santurkar, S., Engstrom, L., Turner, A., and Madry, A.
\newblock Robustness may be at odds with accuracy.
\newblock In \emph{International Conference on Learning Representations}, 2019.

\bibitem[Wong et~al.(2020)Wong, Rice, and Kolter]{wong_fast_2020}
Wong, E., Rice, L., and Kolter, J.~Z.
\newblock Fast is better than free: Revisiting adversarial training.
\newblock In \emph{International Conference on Learning Representations}, 2020.

\bibitem[Wu et~al.(2020)Wu, Wang, Xia, Bailey, and Ma]{wu_skip_2020}
Wu, D., Wang, Y., Xia, S.-T., Bailey, J., and Ma, X.
\newblock Skip connections matter: On the transferability of adversarial
  examples generated with {{ResNets}}.
\newblock In \emph{International Conference on Learning Representations}, 2020.

\bibitem[Xie et~al.(2018)Xie, Wang, Zhang, Ren, and
  Yuille]{xie_mitigating_2018}
Xie, C., Wang, J., Zhang, Z., Ren, Z., and Yuille, A.
\newblock Mitigating adversarial effects through randomization.
\newblock In \emph{International Conference on Learning Representations}, 2018.

\bibitem[Zhang et~al.(2019)Zhang, Yu, Jiao, Xing, Ghaoui, and
  Jordan]{zhang_theoretically_2019}
Zhang, H., Yu, Y., Jiao, J., Xing, E.~P., Ghaoui, L.~E., and Jordan, M.~I.
\newblock Theoretically principled trade-off between robustness and accuracy.
\newblock In \emph{International Conference on Machine Learning}, 2019.

\bibitem[Zhang \& Liang(2019)Zhang and Liang]{zhang_defending_2019}
Zhang, Y. and Liang, P.
\newblock Defending against whitebox adversarial attacks via randomized
  discretization.
\newblock In Chaudhuri, K. and Sugiyama, M. (eds.), \emph{Proceedings of
  Machine Learning Research}, volume~89 of \emph{Proceedings of Machine
  Learning Research}, pp.\  684--693. {PMLR}, April 2019.

\bibitem[Zhao et~al.(2021)Zhao, Liu, and Larson]{zhao_success_2021}
Zhao, Z., Liu, Z., and Larson, M.
\newblock On success and simplicity: A second look at transferable targeted
  attacks.
\newblock \emph{arXiv:2012.11207 [cs]}, February 2021.

\end{thebibliography}

\newpage
\appendix
\onecolumn

\section{Experiment Details} \label{ap:sec:exp_detail}

\subsection{Details on the Image Transformations} \label{ap:ssec:tf_list}

The exact implementation of \rt models and all the transformations will be released.
Here, we provide some details on each of the transformation types and groups.
Then, we describe how we approximate some non-differentiable functions with differentiable ones.

\paragraph{Noise injection}
\begin{itemize}[noitemsep]
    \item \textbf{Erase:} Set the pixels in a box with random size and location to zero. 
    \item \textbf{Gaussian noise:} Add Gaussian noise to each pixel. 
    \item \textbf{Pepper:} Zero out pixels with some probability.
    \item \textbf{Poisson noise:} Add Poisson noise to each pixel.
    \item \textbf{Salt:} Set pixels to one with some probability.
    \item \textbf{Speckle noise:} Add speckle noise to each pixel.
    \item \textbf{Uniform noise:} Add uniform noise to each pixel.
\end{itemize}

\paragraph{Blur filtering}
\begin{itemize}[noitemsep]
    \item \textbf{Box blur:} Blur with randomly sized mean filter.
    \item \textbf{Gaussian blur:} Blur with randomly sized Gaussian filter with randomly chosen variance.
    \item \textbf{Median blur:} Blur with randomly sized median filter.
    \item \textbf{Motion blur:} Blur with kernel for random motion angle and direction.
\end{itemize}

\paragraph{Color-space alteration}
\begin{itemize}[noitemsep]
    \item \textbf{HSV:} Convert to HSV color-space, add uniform noise, then convert back.
    \item \textbf{LAB:} Convert to LAB color-space, add uniform noise, then convert back.
    \item \textbf{Gray scale mix:} Mix channels with random proportions.
    \item \textbf{Gray scale partial mix:} Mix channels with random proportions, then mix gray image with each channel with random proportions.
    \item \textbf{Two channel gray scale mix:} Mix two random channels with random proportions.
    \item \textbf{One channel partial gray:} Mix two random channels with random proportions, then mix gray image with other channel.
    \item \textbf{XYZ:} Convert to XYZ color-space, add uniform noise, then convert back.
    \item \textbf{YUV:} Convert to YUV color-space, add uniform noise, then convert back.
\end{itemize}

\paragraph{Edge detection}
\begin{itemize}[noitemsep]
    \item \textbf{Laplacian:} Apply Laplacian filter.
    \item \textbf{Sobel:} Apply the Sobel operator.
\end{itemize}

\paragraph{Lossy compression}
\begin{itemize}[noitemsep]
    \item \textbf{JPEG compression:} Compress image using JPEG to a random quality.
    \item \textbf{Color precision reduction:} Reduce color precision to a random number of bins.
    \item \textbf{FFT perturbation:} Perform FFT on image and remove each component with some probability.
\end{itemize}

\paragraph{Geometric transforms}
\begin{itemize}[noitemsep]
    \item \textbf{Affine:} Perform random affine transformation on image.
    \item \textbf{Crop:} Crop image randomly and resize to original shape.
    \item \textbf{Horizontal flip:} Flip image across the vertical.
    \item \textbf{Swirl:} Swirl the pixels of an image with random radius and strength.
    \item \textbf{Vertical flip:} Flip image across the horizontal.
\end{itemize}

\paragraph{Stylization}
\begin{itemize}[noitemsep]
    \item \textbf{Color jitter:} Randomly alter the brightness, contrast, and saturation.
    \item \textbf{Gamma:} Randomly alter gamma.
    \item \textbf{Sharpen:} Apply sharpness filter with random strength.
    \item \textbf{Solarize:} Solarize the image.
\end{itemize}

\paragraph{Non-differentiable (for BPDA Tests Only)}
\begin{itemize}[noitemsep]
    \item \textbf{Adaptive histogram:} Equalize histogram in patches of random kernel size.
    \item \textbf{Chambolle denoise:} Apply Chambolle's total variation denoising algorithm with random weight (can be implemented differentiably but was not due to time constraints).
    \item \textbf{Contrast stretching:} Pick a random minimum and maximum pixel value to rescale intensities (can be implemented differentiably but was not due to time constraints).
    \item \textbf{Histogram:} Equalize histogram using a random number of bins.
\end{itemize}

\paragraph{Unused transforms from BaRT}
\begin{itemize}[noitemsep]
    \item \textbf{Seam carving:} Algorithm used in \citet{raff_barrage_2019} has been patented and is no longer available for open-source use.
    \item \textbf{Wavelet denoising:} The implementation in \citet{raff_barrage_2019} is incomplete. 
    \item \textbf{Salt \& pepper:} We have already used salt and pepper noise separately.
    \item \textbf{Non-local means denoising:} The implementation of NL means denoising in \citet{raff_barrage_2019} is too slow.
\end{itemize}

\subsection{Experiment Details} \label{ap:ssec:exp_setup}

All of the experiments are evaluated on 1000 randomly chosen test samples. 
Since we choose the default $n$ to be 20 for inference and 10 for the attacks, the experiments are at least 10 times more expensive than usual, and we cannot afford enough computation to run a large number of experiments on the entire test set.
The networks used in this paper are ResNet-34~\cite{he_deep_2016} for Imagenette and Pre-activation ResNet-20~\cite{he_identity_2016} for CIFAR-10. 
In all of the experiments, we use a learning rate of 0.05,
batch size of 128, and weight decay of 0.0005. 
We use cosine annealing schedule~\cite{loshchilov_sgdr_2017} for the learning rate with a period of
10 epochs which also doubles after every period. 
All models are trained for 70 epochs, and we save the weights with the highest accuracy on the held-out validation data (which does not overlap with the training or test set).
For adversarially trained \rt defenses, the cosine annealing step is set to 10 and the training lasts for 70 epochs to reduce the computation.
To help the training converge faster, we pre-train these \rt models on clean data before turning on adversarial training as suggested by \citet{gupta_improving_2020}.

\subsection{Details on BPDA Experiments} \label{ap:ssec:bpda_detail}

\begin{figure}[t!]
    \centering
    \includegraphics[width=\linewidth]{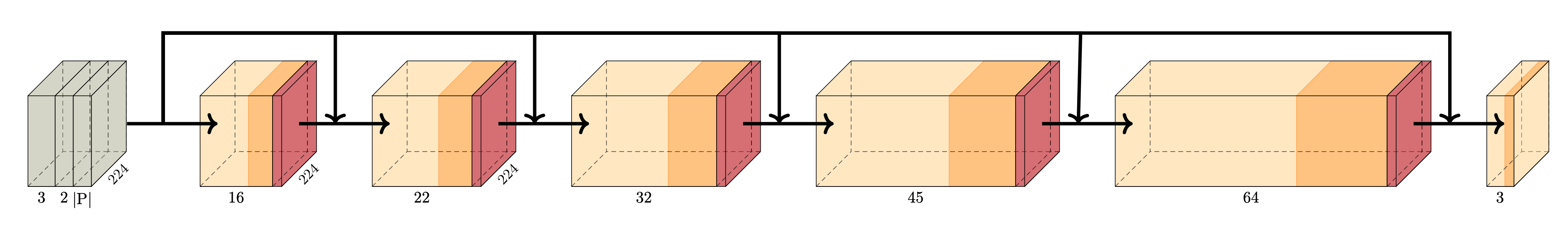}
    \vspace{-5pt}
    \caption{Fully-convolutional BPDA network from \citet{raff_barrage_2019}. The network has six convolutional layers. 
    All layers have a stride of 1. The first five layers have kernel size of 5 and padding size of 2, and the last layer has a kernel size of 3 and padding size of 1. 
    The input consists of more than 5 channels, 3 of which are for the image RGB channels, 2 of which are CoordConv channels that include the coordinates of each pixel at that pixel's location, and the remaining channels are the parameters for the transformation copied at each pixel location. The network contains a skip connection from the input to each layer except the final layer.}
    \label{fig:bpda}
    \vspace{-5pt}
\end{figure}

We used the following setup for the differentiability related experiments conducted in Section \ref{ssec:bpda-exp}:
\begin{itemize}[noitemsep]
    \item Each accuracy is an average over 10 trials on the same set of 1000 Imagenette images.
    \item The defense samples $S = 10$ transforms from the full set of $K$ transforms.
    \item The image classifier uses a ResNet-50 architecture like in \citet{raff_barrage_2019} trained on transformed images for $30$ epochs.
    \item The attack uses $40$ PGD steps of size $4/255$ with an $\epsilon=16/255$ to minimize the EoT objective.
\end{itemize}
 
\begin{figure*}
    \centering
    \begin{subfigure}[b]{\linewidth}
    \centering
        \includegraphics[trim=2cm 0.75cm 6cm 1cm,clip,width=0.49\linewidth]{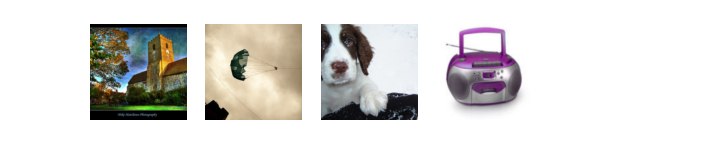}
        \caption{Original}
        \vspace{10pt}
    \end{subfigure}
    \newline
    \begin{subfigure}[b]{0.49\linewidth}
        \centering
        \includegraphics[trim=2cm 0.75cm 6cm 1cm,clip,width=\linewidth]{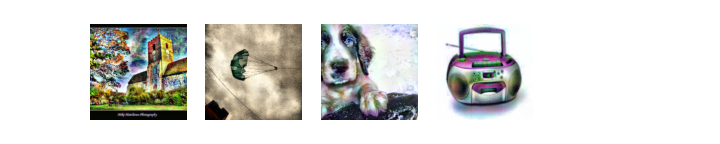}
        \newline
        \includegraphics[trim=2cm 0.75cm 6cm 1cm,clip,width=\linewidth]{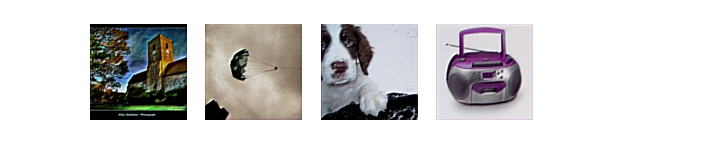}
        \caption{Adaptive histogram}
        \vspace{10pt}
    \end{subfigure}
    \hfill
    \begin{subfigure}[b]{0.49\linewidth}
        \centering
        \includegraphics[trim=2cm 0.75cm 6cm 1cm,clip,width=\linewidth]{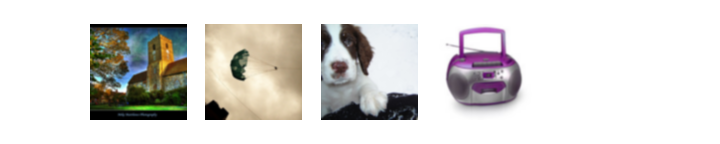}
        \newline
        \includegraphics[trim=2cm 0.75cm 6cm 1cm,clip,width=\linewidth]{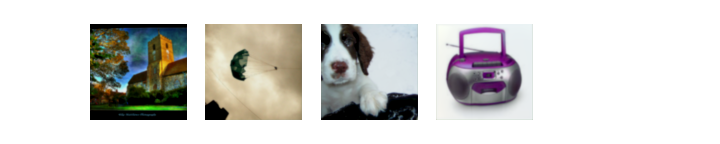}
        \caption{Box blur}
        \vspace{10pt}
    \end{subfigure}
    \newline
    \begin{subfigure}[b]{0.49\linewidth}
        \centering
        \includegraphics[trim=2cm 0.75cm 6cm 1cm,clip,width=\linewidth]{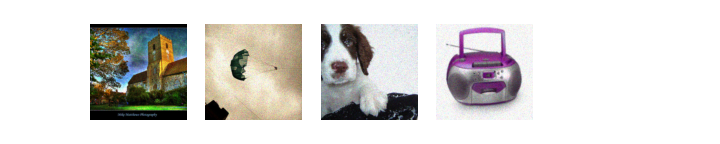}
        \newline
        \includegraphics[trim=2cm 0.75cm 6cm 1cm,clip,width=\linewidth]{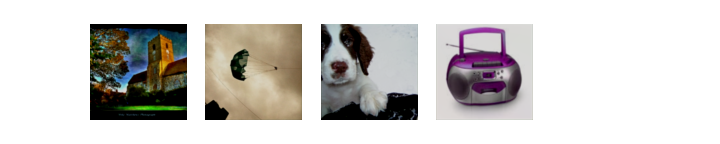}
        \caption{Poisson noise}
        \vspace{10pt}
    \end{subfigure}
    \hfill
    \begin{subfigure}[b]{0.49\linewidth}
        \centering
        \includegraphics[trim=2cm 0.75cm 6cm 1cm,clip,width=\linewidth]{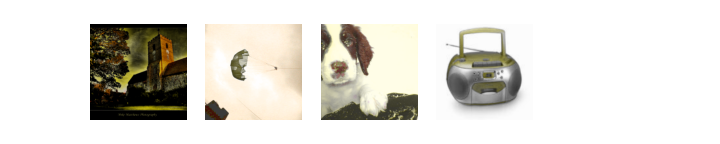}
        \newline
        \includegraphics[trim=2cm 0.75cm 6cm 1cm,clip,width=\linewidth]{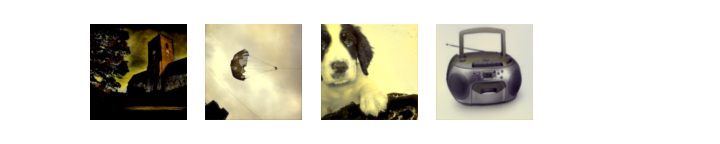}
        \caption{HSV color alteration}
        \vspace{10pt}
    \end{subfigure}
    \newline
    \begin{subfigure}[b]{0.49\linewidth}
        \centering
        \includegraphics[trim=2cm 0.75cm 6cm 1cm,clip,width=\linewidth]{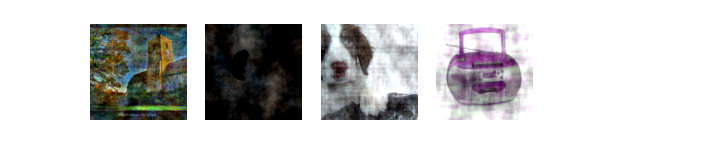}
        \newline
        \includegraphics[trim=2cm 0.75cm 6cm 1cm,clip,width=\linewidth]{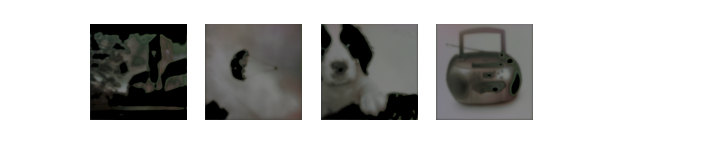}
        \caption{FFT}
    \end{subfigure}
    \hfill
    \begin{subfigure}[b]{0.49\linewidth}
        \centering
        \includegraphics[trim=2cm 0.75cm 6cm 1cm,clip,width=\linewidth]{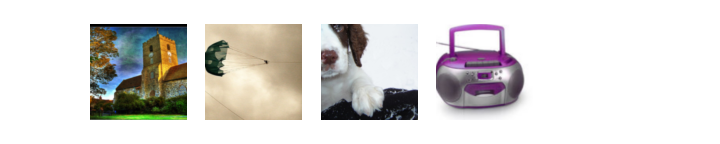}
        \newline
        \includegraphics[trim=2cm 0.75cm 6cm 1cm,clip,width=\linewidth]{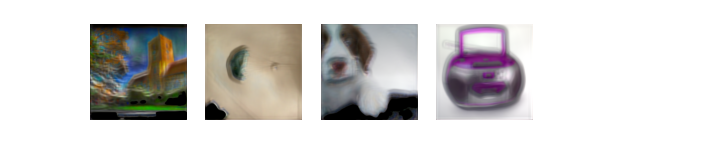}
        \caption{Crop}
    \end{subfigure}
    \caption{Comparison of the true transformed outputs (top row) and outputs of respective BPDA networks (bottom row) for six different transformation types.}
    \label{fig:bpda_comparison}
\end{figure*}

The BPDA network architecture is the same used by \citet{raff_barrage_2019} and is outlined in \figref{fig:bpda}.
All BPDA networks were trained using Adam with a learning rate of $0.01$ for 10 epochs.
All networks achieve a per-pixel MSE below $0.01$. 
The outputs of the BPDA networks are compared to the true transform outputs for several different transform types in \figref{fig:bpda_comparison}.
The specific set of transforms used in each defense are the following:
\begin{itemize}
    \item \textbf{BaRT (all):} adaptive histogram, histogram, bilateral blur, box blur, Gaussian blur, median blur, contrast stretching, FFT, gray scale mix, gray scale partial mix, two channel gray scale mix, one channel gray scale mix, HSV, LAB, XYZ, YUV, JPEG compression, Gaussian noise, Poisson noise, salt, pepper,  color precision reduction, swirl, Chambolle denoising,  crop.
    \item \textbf{BaRT (only differentiable):} all of the BaRT all transforms excluding adaptive histogram, histogram, contrast stretching, and Chambolle denoising.
\end{itemize}

\section{Details of the Attacks} \label{ap:sec:attack}

\subsection{Differentiable Approximation}

Some of the transformations contain non-differentiable operations which can be easily approximated with differentiable functions.
Specifically, we approximate the rounding function in JPEG compression and color precision reduction, and the modulo operator in all transformations that require conversion between RGB and HSV color-spaces (HSV alteration and color jitter).
Note that we are not using the non-differentiable transform on the forward pass and a differentiable approximation on the backward pass (like in BPDA).
Instead, we are using the differentiable version both when performing the forward pass and when computing the gradient.

We take the approximation of the rounding function from \citet{shin_jpegresistant_2017} shown in \eqref{eq:diff_round}.
\begin{align} \label{eq:diff_round}
    \lfloor x \rceil_\text{approx} = \lfloor x \rceil + (x - \lfloor x \rceil)^3
\end{align}
For the modulo or the remainder function, we approximate it using the above differentiable rounding function as a basis.
\begin{align} \label{eq:diff_mod}
    \mathrm{mod}(x) &= \begin{cases}
    x - \lfloor x \rceil \qquad\quad\mathrm{if}~x > \lfloor x \rceil \\
    x - \lfloor x \rceil + 1 \quad~\mathrm{otherwise}
    \end{cases}
\end{align}
To obtain a differentiable approximation, we can replace the rounding operator with its smooth version in \eqref{eq:diff_round}.
This function (approximately) returns decimal numbers or a fractional part of a given real number, and it can be scaled to approximate a modulo operator with any divisor.

Note that these operators are step functions and are differentiable almost everywhere, like ReLU.
However, their derivatives are always zero (unlike ReLU), and so a first-order optimization algorithm would still fail on these functions. 

\subsection{Effect of the Permutation of the Transformations} \label{ap:ssec:tf-perm}

We mentioned in Section~\ref{ssec:tf_params} that a permutation of the transforms $\{\tau^{(s)}\}_{s=1}^S$ is randomly sampled for each of the $n$ samples. 
However, we found that in practice, this leads to high-variance estimates of the gradients.
On the other hand, fixing the permutation across $n$ samples in each attack iteration (i.e., $\tau$ is fixed but not $\alpha$ or $\beta$) results in lower variance and hence, a stronger attack, even though the gradient estimates are biased as $\tau$ is fixed. 
For instance, with fixed permutation, adversarial accuracy achieved by EoT attack is $51.44$ where the baseline EoT with completely random permutation is $70.79$.
The variance also reduces from $0.97$ to $0.94$.

Additionally, the fixed permutation reduces the computation time as all transformations can be applied in batch. All of the attacks reported in this paper, apart from the baseline, use this fixed permutation.

\begin{table*}[t!]
\small
\centering
\caption{Comparison of different attack techniques on our best \rt model. Lower means stronger attack. This table only shows the numerical results plotted in Fig.~\ref{fig:attack_loss_ens}.}
\label{tab:main_attack}
\begin{tabular}{@{}lrrrrrr@{}}
\toprule
\multirow{2}{*}{Attacks} & \multicolumn{3}{c}{Adv. acc. with varying attack steps ($n=10$)} & \multicolumn{3}{c}{Adv. acc. with varying $n$ (attack steps = 200)} \\ \cmidrule(l){2-4} \cmidrule(l){5-7}
                         & $50$  & $200$ & $800$ & $5$  & $10$ & $20$ \\ \midrule
Baseline       & $82.34 \pm 0.43$ & $73.36 \pm 0.37$ & $71.70 \pm 0.39$ & $74.81 \pm 0.47$ & $74.46 \pm 0.55$ & $76.06 \pm 0.29$ \\
CE (softmax)             & $82.37 \pm 0.39$ & $71.05 \pm 0.36$ & $65.06 \pm 0.39$ & $73.82 \pm 0.35$ & $70.71 \pm 0.53$ & $68.51 \pm 0.33$ \\
Linear (logits)                  & $80.67 \pm 0.50$ & $66.11 \pm 0.58$ & $58.26 \pm 0.62$ & $70.67 \pm 0.41$ & $66.59 \pm 0.57$ & $62.48 \pm 0.41$ \\ \midrule
Linear+MB              & $\bm{78.51} \pm 0.45$ & $72.66 \pm 0.50$ & $65.28 \pm 0.41$ & $72.47 \pm 0.39$ & $72.51 \pm 0.55$ & $71.06 \pm 0.32$ \\
Linear+LinBP           & $82.90 \pm 0.50$ & $70.57 \pm 0.32$ & $65.15 \pm 0.43$ & $75.24 \pm 0.35$ & $72.73 \pm 0.40$ & $70.02 \pm 0.31$ \\
Linear+SGM             & $80.10 \pm 0.43$ & $\bm{63.75} \pm 0.21$ & $\bm{51.68} \pm 0.35$ & $\bm{66.93} \pm 0.43$ & $\bm{62.57} \pm 0.31$ & $59.61 \pm 0.55$\\
Linear+TG              & $80.78 \pm 0.56$ & $68.70 \pm 0.34$ & $\bm{59.69} \pm 0.57$ & $71.72 \pm 0.41$ & $67.84 \pm 0.50$ & $65.63 \pm 0.50$ \\ 
\bottomrule
\end{tabular}
\end{table*}

\begin{figure}
     \centering
     \hfill
     \begin{subfigure}[b]{0.4\textwidth}
         \centering
         \includegraphics[width=\textwidth]{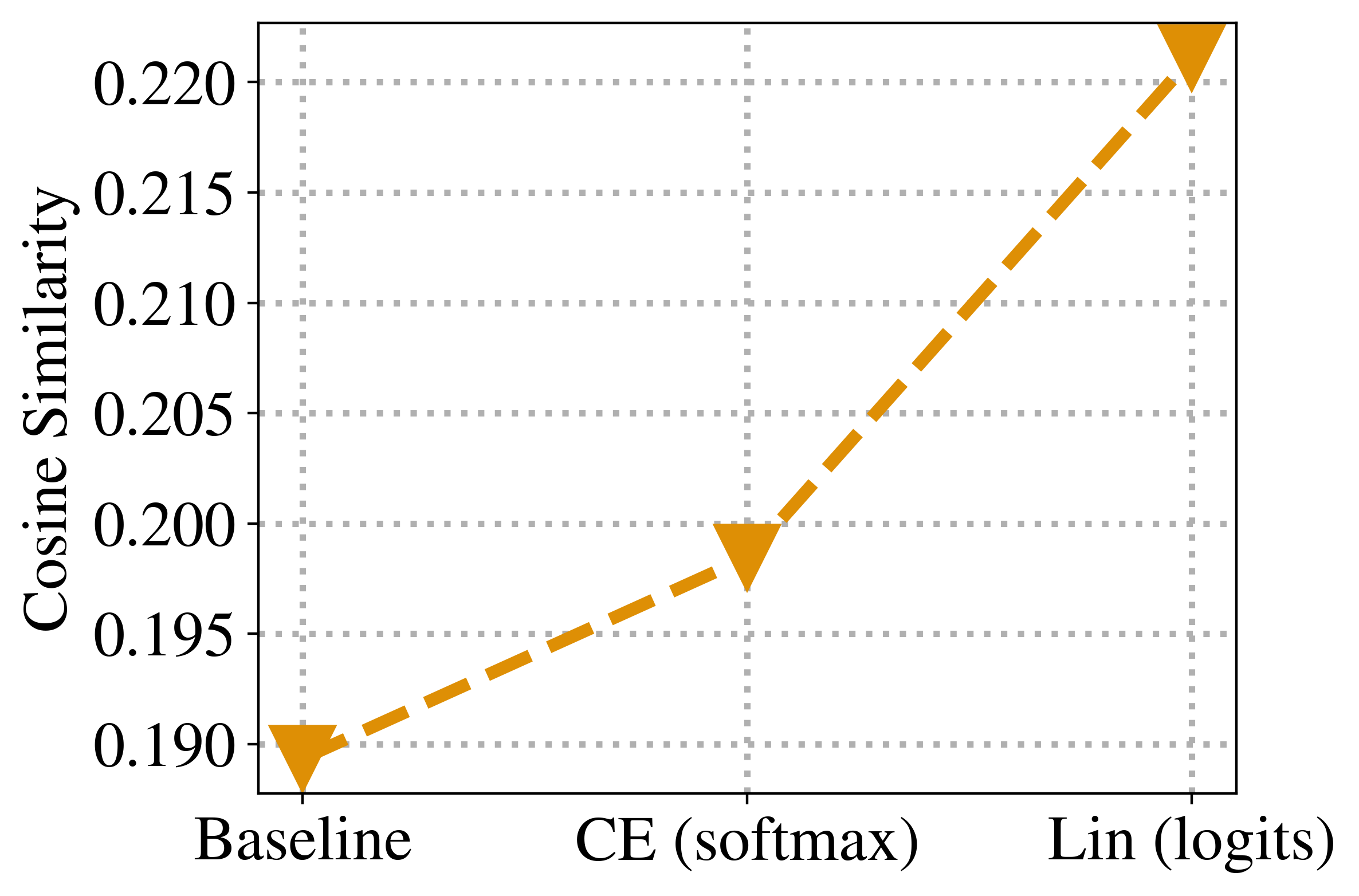}
         \caption{Cosine Similarity}
         \label{fig:loss_var_1}
     \end{subfigure}
     \hfill
     \begin{subfigure}[b]{0.4\textwidth}
         \centering
         \includegraphics[width=\textwidth]{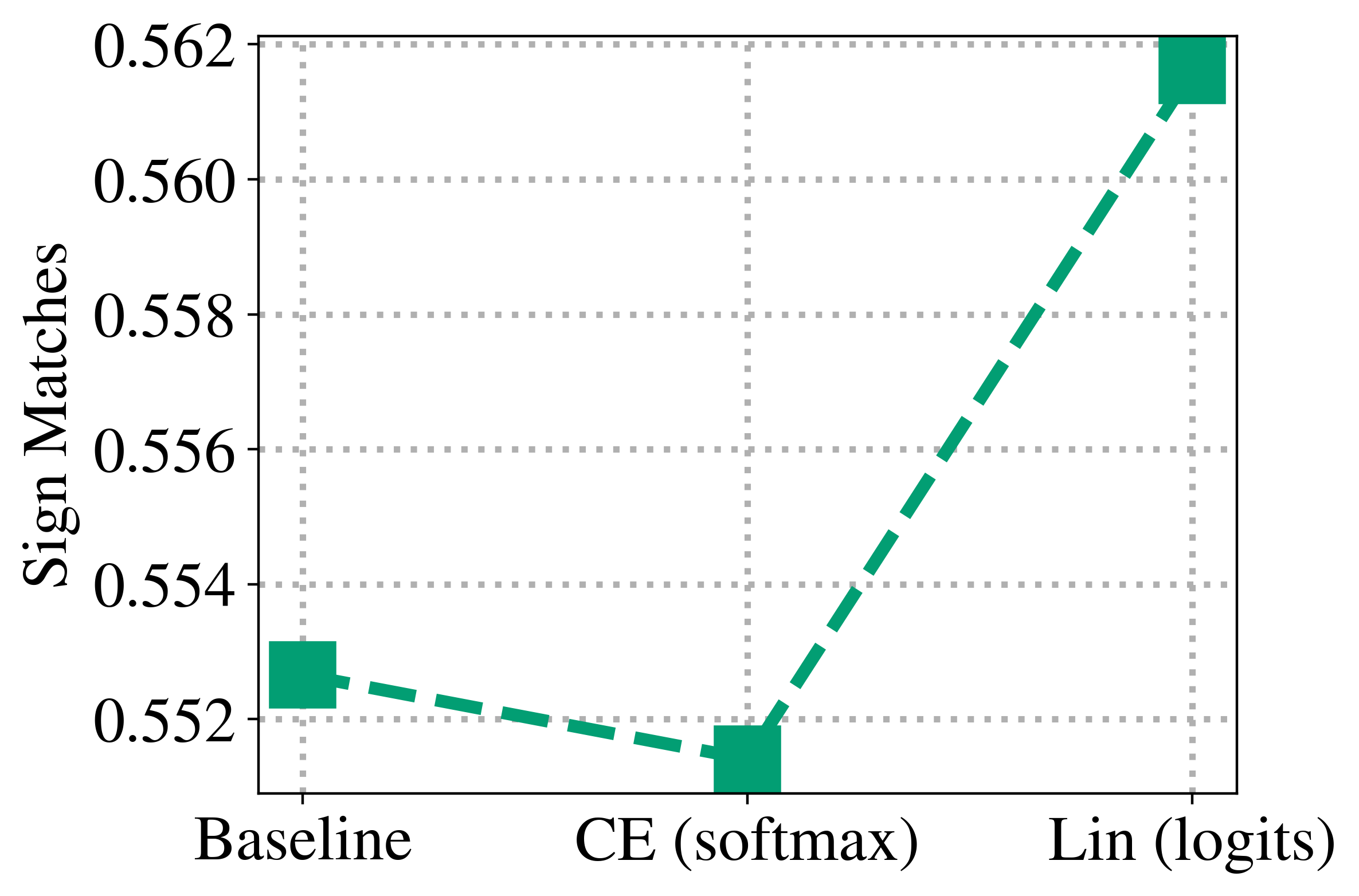}
         \caption{Sign Matches}
         \label{fig:loss_var_2}
     \end{subfigure}
     \hfill
     \phantom{.}
     \caption{(a) Cosine similarity and (b) percentage of sign matches for three pairs of attack loss functions and decision rules: CE loss with EoT ``Baseline'', CE loss on mean softmax probability ``CE (softmax)'', and linear loss on logits ``Lin (logits)''.}
     \label{fig:loss_var}
\end{figure}

\begin{figure}
     \centering
     \hfill
     \begin{subfigure}[b]{0.4\textwidth}
         \centering
         \includegraphics[width=\textwidth]{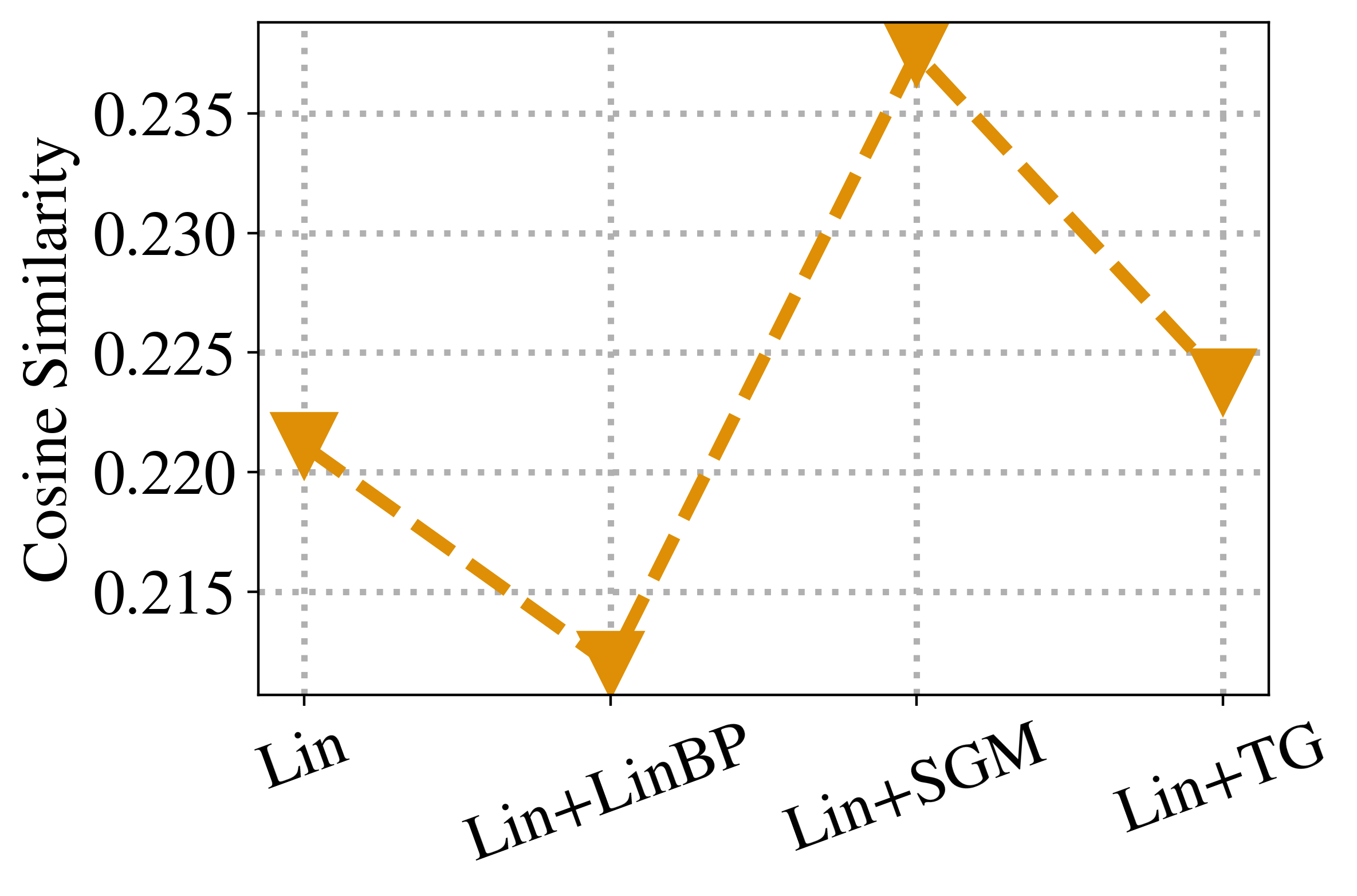}
         \caption{Cosine Similarity}
         \label{fig:ens_var_1}
     \end{subfigure}
     \hfill
     \begin{subfigure}[b]{0.4\textwidth}
         \centering
         \includegraphics[width=\textwidth]{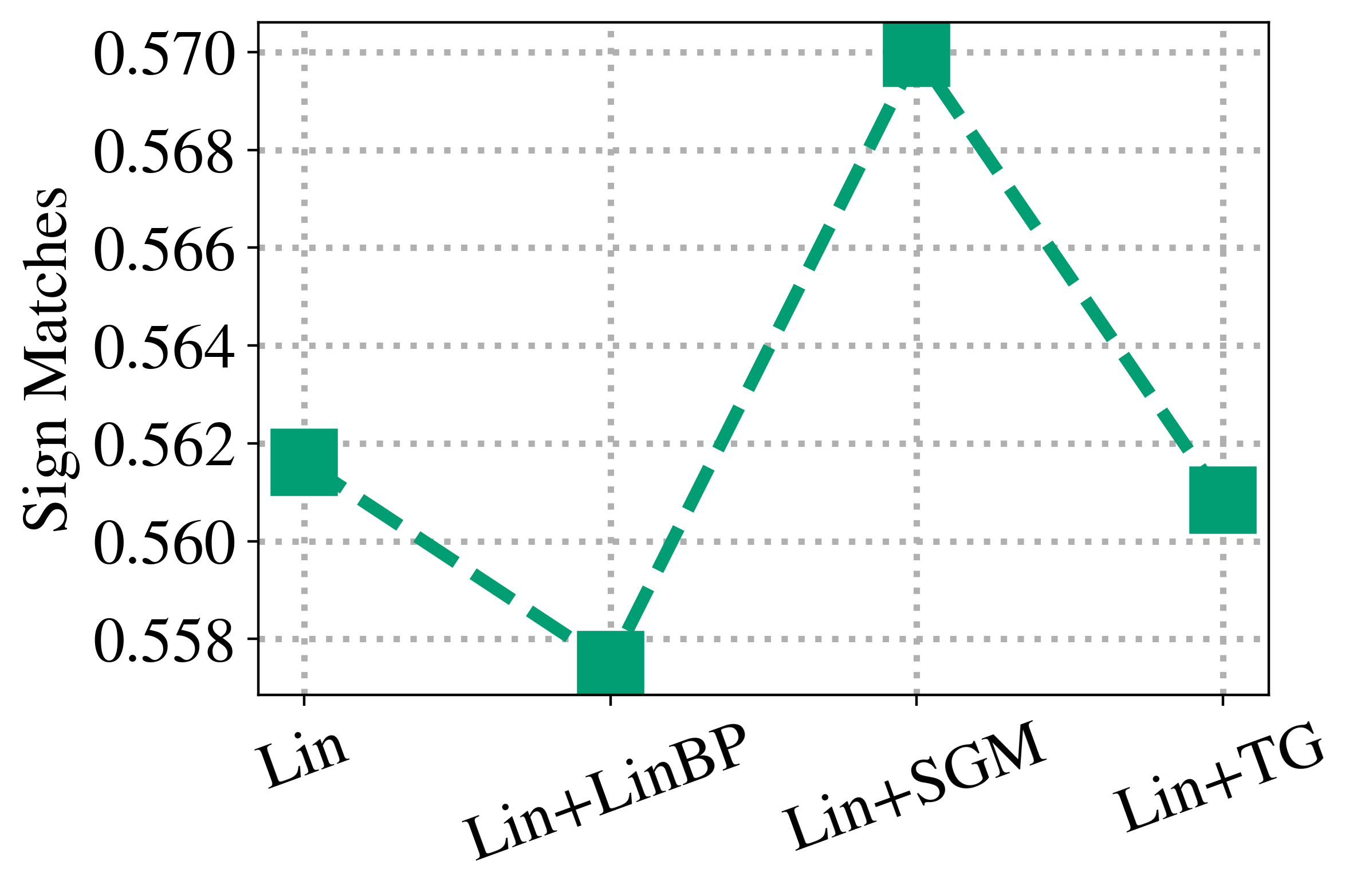}
         \caption{Sign Matches}
         \label{fig:ens_var_2}
     \end{subfigure}
     \hfill
     \phantom{.}
     \caption{(a) Cosine similarity and (b) percentage of sign matches for the linear loss and its combinations with three transfer attack techniques: Linear Backward Pass ``LinBP'', Skip Gradient Method ``SGM'', and targeted ``TG''.}
     \label{fig:ens_var}
\end{figure}

\subsection{Variance of Gradients} \label{ap:ssec:grad_var}

We have described how we compute the sample variance of the gradients in Section~\ref{ssec:var_sgd}.
Here, we provide detailed calculations of the other three metrics.
First, the unbiased variance is computed as normal with an additional normalization by dimension.
\begin{align}
    \mu_{n} &\coloneqq \frac{1}{n} \sum_{j=1}^n \nabla \hat{G}_{1,j} \label{eq:mean_grad} \\
    \sigma_{n}^2 &\coloneqq \frac{1}{d}\frac{1}{n-1} \sum_{j=1}^n \norm{\mu_{n} - \hat{G}_{1,j}}_2^2 \label{eq:var_grad}
\end{align}
where $\hat{G}_1$ is the signed gradients where the loss is estimated with one sample as defined in Algorithm~\ref{alg:attack}.

The cosine similarity is computed between the mean gradient and all $n$ samples and then averaged.
\begin{align}
    \text{cos}_{n} \coloneqq \frac{1}{n} \sum_{j=1}^n \frac{\inner{\hat{G}_{1,j}, \mu_{n}}}{\norm{\hat{G}_{1,j}}_2 \cdot \norm{\mu_{n}}_2}
\end{align}
Lastly, the sign matching percentage is 
\begin{align}
    \text{sign\_match}_{n}. \coloneqq \frac{1}{n} \sum_{j=1}^n \frac{1}{d} \sum_{i=1}^d \mathbbm{1}\{[\hat{G}_{1,j}]_i = [\mu_{n}]_i\}
\end{align}

\figref{fig:loss_var} and \figref{fig:ens_var} plot the cosine similarly and the sign matching for varying loss functions and varying transfer attacks, respectively. 
Similarly to \figref{fig:main_var}, better attacks result in less spread of the gradient samples which corresponds to higher cosine similarity and sign matching percentage.

\begin{figure}[t!]
\centering
    \hfill
    \begin{subfigure}[b]{0.4\textwidth}
        \centering
        \includegraphics[width=\textwidth]{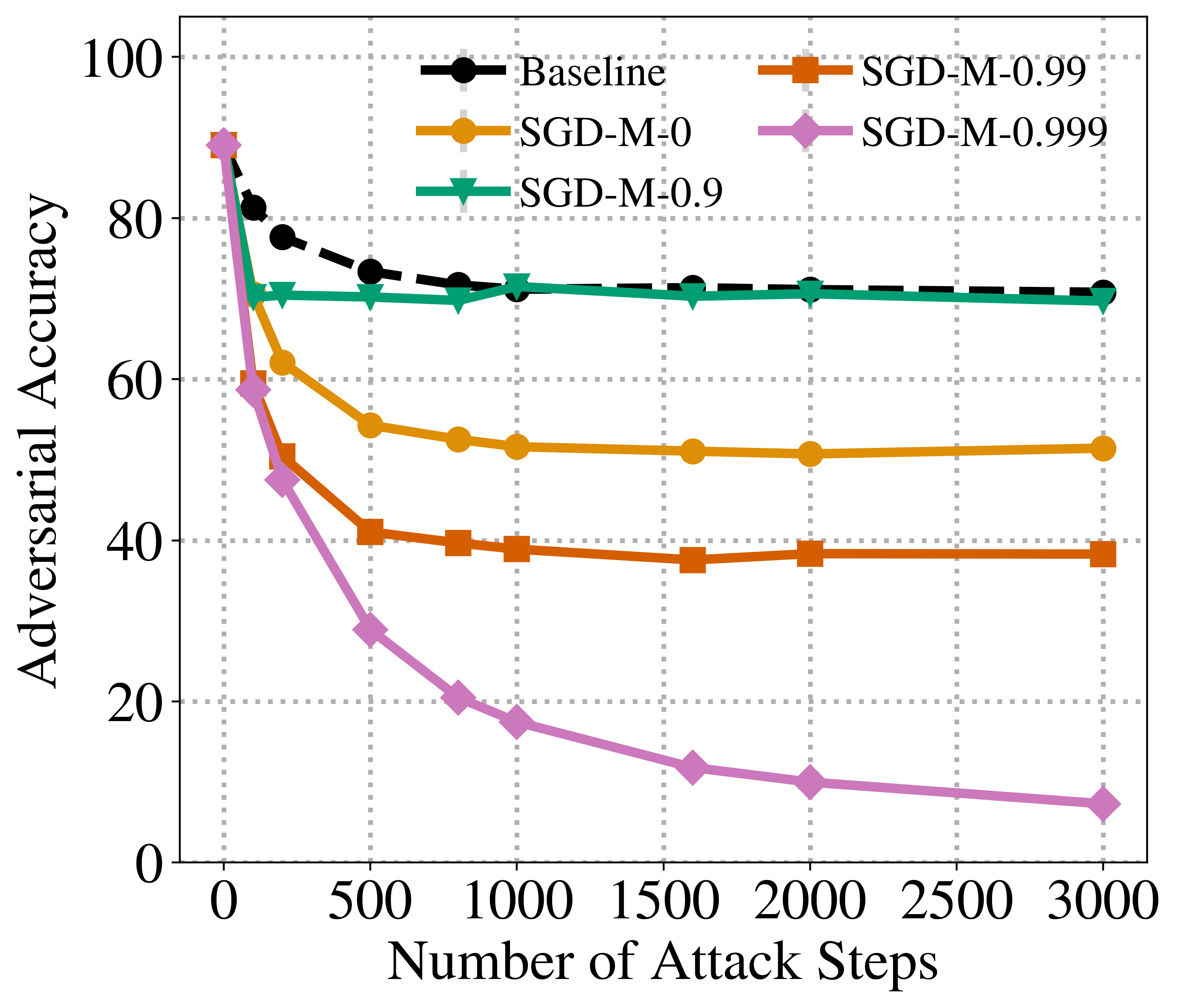}
        \caption{SGD with varying momentum constants}
        \label{fig:atk_img_rand_sgd}
    \end{subfigure}
    \hfill
    \begin{subfigure}[b]{0.4\textwidth}
        \centering
        \includegraphics[width=\textwidth]{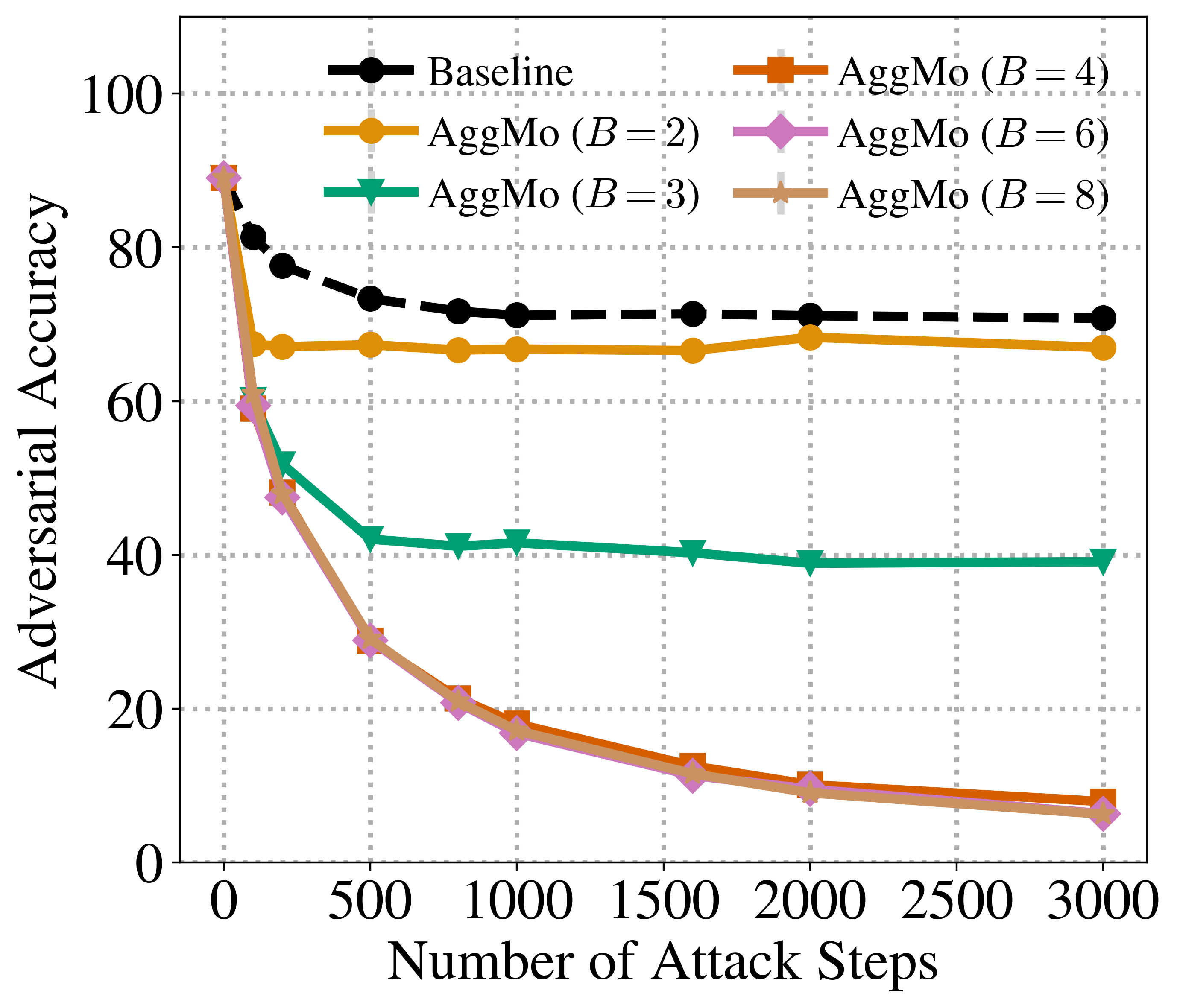}
        \caption{AggMo with varying $B$'s}
        \label{fig:atk_img_rand_aggmo}
    \end{subfigure}
    \hfill\phantom{x}
    \caption{Effectiveness of the optimizers, (a) SGD and (b) AggMo, with varying momentum parameters. Increasing $B$ for AggMo in this case monotonically reduces the final adversarial accuracy until $B=4$ where it plateaus. This is more predictable and stable than increasing the momentum constant in SGD.}
    \label{fig:atk_img_rand_opt}
\end{figure}
\section{Details on Bayesian Optimization} \label{ap:sec:bayes}

\begin{algorithm}[tb]
\caption{Tuning and training \rt defense.}
\label{alg:bo}
\begin{algorithmic}
    \STATE {\bfseries Input:} Set of transformation types, $n$, $p$, $\epsilon$
    \STATE {\bfseries Output:} $g^*(\cdot), \mathcal{R}, \mathcal{R}_{p,\epsilon}$
    \STATE {\bfseries Data:} Training data $\left(\bm{X}^{\mathrm{train}}, \bm{Y}^{\mathrm{train}}\right)$, test data $\left(\bm{X}^{\mathrm{test}}, \bm{Y}^{\mathrm{test}}\right)$
    \STATE \textcolor{blue}{\texttt{// Starting Bayesian optimization (BO)}}
    \STATE Sub-sample $\left(\bm{X}^{\mathrm{train}}, \bm{Y}^{\mathrm{train}}\right)$ and split it into BO's training data $\left(\bm{X}^{\mathrm{train}}_{\mathrm{BO}}, \bm{Y}^{\mathrm{train}}_{\mathrm{BO}}\right)$ and validation data $\left(\bm{X}^{\mathrm{val}}_{\mathrm{BO}}, \bm{Y}^{\mathrm{val}}_{\mathrm{BO}}\right)$. \label{alg:line:subsample}
    \STATE $\mathcal{R}_{p,\epsilon}^* \gets 0$ \hfill\textcolor{blue}{\texttt{// Best adversarial accuracy}}
    \STATE $\{(p^*_i, \alpha^*_i)\}_{i=1}^{K} \gets 0$ \hfill\textcolor{blue}{\texttt{// Best \rt hyperparameters}}
    \FOR{$\mathrm{step}=1$ {\bfseries to} MAX\_BO\_STEPS}
        \STATE \textcolor{blue}{\texttt{// Running one trial of BO}}
        \STATE BO specifies $\{(p_i, \alpha_i)\}_{i=1}^{K}$ to evaluate.
        \STATE Train an \rt model on $\left(\bm{X}^{\mathrm{train}}_{\mathrm{BO}}, \bm{Y}^{\mathrm{train}}_{\mathrm{BO}}\right)$ with hyperparameters $\{(p_i, \alpha_i)\}_{i=1}^{K}$ to obtain $g$.
        \STATE Test $g$ by computing $\mathcal{R}_{p,\epsilon}$ on $\left(\bm{X}^{\mathrm{val}}_{\mathrm{BO}}, \bm{Y}^{\mathrm{val}}_{\mathrm{BO}}\right)$ using a weak but fast attack. \label{alg:line:test}
        \IF{$\mathcal{R}_{p,\epsilon} > \mathcal{R}_{p,\epsilon}^*$}
            \STATE $\mathcal{R}_{p,\epsilon}^* \gets \mathcal{R}_{p,\epsilon}$
            \STATE $\{(p^*_i, \alpha^*_i)\}_{i=1}^{K} \gets \{(p_i, \alpha_i)\}_{i=1}^{K}$
        \ELSIF{No improvement for some steps}
            \STATE break
        \ENDIF
    \ENDFOR
    \STATE \textcolor{blue}{\texttt{// Full training of \rt}}
    \STATE  Train an \rt model on $\left(\bm{X}^{\mathrm{train}}, \bm{Y}^{\mathrm{train}}\right)$ with best hyperparameters $\{(p^*_i, \alpha^*_i)\}_{i=1}^{K}$ to obtain $g^*$. \label{alg:line:full_train}
    \STATE Evaluate $g^*$ by computing $\mathcal{R}$ and $\mathcal{R}_{p,\epsilon}$ on $\left(\bm{X}^{\mathrm{test}}, \bm{Y}^{\mathrm{test}}\right)$ using a strong attack. \label{alg:line:full_test}
\end{algorithmic}
\end{algorithm}

One major challenge in implementing an \rt defense is selecting the defense hyperparameters which include the $K$ transformation types, the number of transformations to apply ($S$), and their parameters ($a$ and $p$).
To improve the robustness of \rt defense, we use Bayesian optimization (BO), a well-known black-box optimization technique, to fine-tune $a$ and $p$~\citep{snoek_practical_2012}.
In this case, BO models the hyperparameter tuning as a Gaussian process where the objective function takes in $a$ and $p$, trains a neural network as a backbone for an \rt defense, and outputs adversarial accuracy under some pre-defined $\ell_\infty$-budget $\epsilon$ as the metric used for optimization.

Since BO quickly becomes ineffective as we increase the dimensions of the search space, we choose to tune either $a$ or $p$, never both, for each of the $K$ transformation types.
For transformations that have a tunable $a$, we fix $p = 1$ (e.g., noise injection, affine transform). 
For the transformations without an adjustable strength $a$, we only tune $p$ (e.g., Laplacian filter, horizontal flip).
Additionally, because BO does not natively support categorical or integral variables, we experiment with different choices for $K$ and $S$ without the use of BO.
Therefore, our BO problem must optimize over $K$ (up to $33$) variables, far more than are typically present when doing model hyperparamter tuning using BO.

Mathematically, the objective function $\psi$ is defined as
\begin{align}
    \psi : [0, 1]^K \to \mathcal{R}_{\infty,\epsilon} \in [0, 1]
\end{align}
where the input is $K$ real numbers between $0$ and $1$, and $\mathcal{R}_{\infty,\epsilon}$ denotes the adversarial accuracy or the accuracy on $x_{\mathrm{adv}}$ as defined in \eqref{eq:adv}.
Since $\psi$ is very expensive to evaluate as it involves training and testing a large neural network, we employ the following strategies to reduce the computation: (1) only a subset of the training and validation set is used, (2) the network is trained for fewer epochs with a cosine annealing learning rate schedule to speed up convergence~\cite{loshchilov_sgdr_2017}, and (3) the attack used for computing $\mathcal{R}_{\infty,\epsilon}$ is weaker but faster.
Even with these speedups, one BO run still takes approximately two days to complete on two GPUs (Nvidia GeForce GTX 1080 Ti).
We also experimented with other sophisticated hyperparameter-tuning algorithms based on Gaussian processes~\cite{bergstra_making_2013,kandasamy_tuning_2020,falkner_bohb_2018} but do not find them more effective.
We summarize the main steps for tuning and training an \rt defense in Algorithm~\ref{alg:bo}.

We use the Ray Tune library for \rt's hyperparameter tuning in Python~\cite{liaw_tune_2018}.
The Bayesian optimization tool is implemented by \citet{nogueira_bayesian_2014}, following analyses and instructions by \citet{snoek_practical_2012} and \citet{brochu_tutorial_2010}.
As mentioned in Section~\ref{sec:bayesopt}, we sub-sample the data to reduce computation for each BO trial.
Specifically, we use 20\% and 10\% of the training samples for Imagenette and CIFAR-10 respectively (Algorithm~\ref{alg:bo}, line~\ref{alg:line:subsample}) as Imagenette has a much smaller number of samples in total.
The models are trained with the same transformations and hyperparameters used during inference, and here, $n$ is set to 1 during training, just as is done during standard data augmentation.
We use 200 samples to evaluate each BO run in line~\ref{alg:line:test} of Algorithm~\ref{alg:bo} with only 100 steps and $n=10$.

One BO experiment executes two BO's in parallel. The maximum number of BO runs is 160, but we terminate the experiment if no improvement has been made in the last 40 runs after a minimum of 80 runs have taken place.
The runtime depends on $S$ and the transformation types used.
In our typical case, when all 33 transformation types are used and $S=14$, one BO run takes almost an hour on an Nvidia GeForce GTX 1080 Ti for Imagenette.
One BO experiment then takes about two days to finish.

In line~\ref{alg:line:full_train} and \ref{alg:line:full_test} of Algorithm~\ref{alg:bo}, we now use the full training set and 1000 test samples as mentioned earlier. 
During the full training, $n$ is set to four which increases the training time by approximately four times.
We find that using a larger $n$ is beneficial to both the clean and the adversarial accuracy, but $n$ larger than four does not make any significant difference.

\subsection{Details on the Final \rt Model} \label{ap:ssec:final}

We run multiple BO experiments (Algorithm~\ref{alg:bo}) on different subsets of transformation types to identify which transformations are most/least effective in order to reduce $K$ as well as the number of hyperparameters our final run of BO has to tune.
We then repeat Algorithm~\ref{alg:bo} initialized with the input-output pairs from the prior runs of BO to obtain a new set of hyperparameters.
Finally, we remove the transformations whose $p$ or $a$ has been set to zero by the first run of BO, and we run BO once more with this filtered subset of transformations.
At the end of this expensive procedure, we obtain the best and final \rt model that we use in the experiments throughout this paper.
For Imagenette, the final set of 18 transformation types used in this model are color jitter, erase, gamma, affine, horizontal flip, vertical flip, Laplacian filter, Sobel filter, Gaussian blur, median blur, motion blur, Poisson noise, FFT, JPEG compression, color precision reduction, salt noise, sharpen, and solarize.
$S$ is set to 14.

\section{Additional Experiments on the \rt Model} \label{ap:sec:defense}

\subsection{Decision Rules and Number of Samples} \label{ap:ssec:rule}

\begin{figure}[t!]
    \centering
    \includegraphics[width=0.4\textwidth]{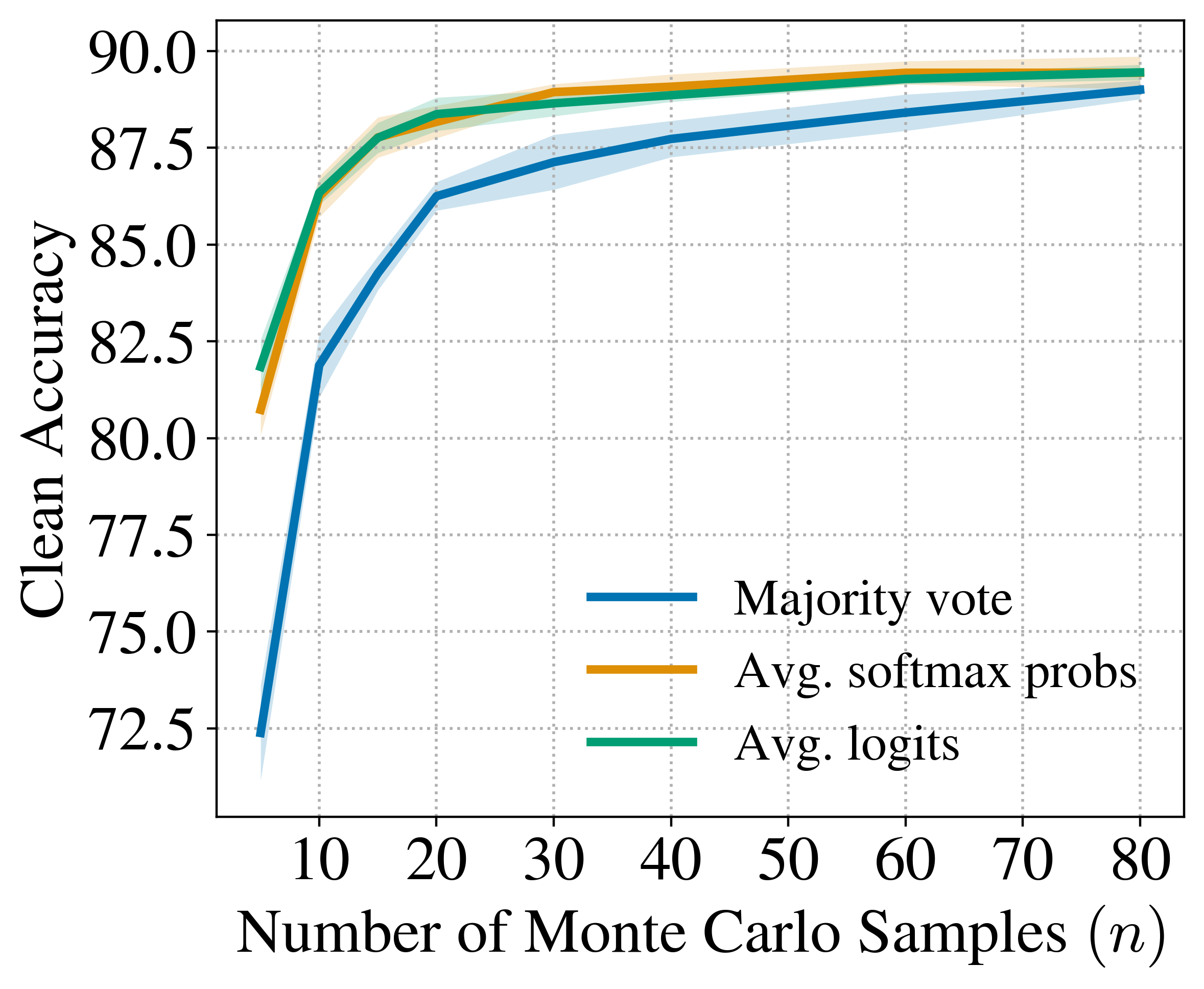}
    \caption{Clean accuracy of our best \rt model computed with three decision rules for obtaining the final prediction from the $n$ output samples. The rules are majority vote (red), average softmax probability (blue), and average logits (green). The shaded areas represent the 95\% confidence interval for each decision rule.}
    \label{fig:clean_rule}
\end{figure}

\begin{figure}[t!]
    \centering
    \includegraphics[width=0.4\textwidth]{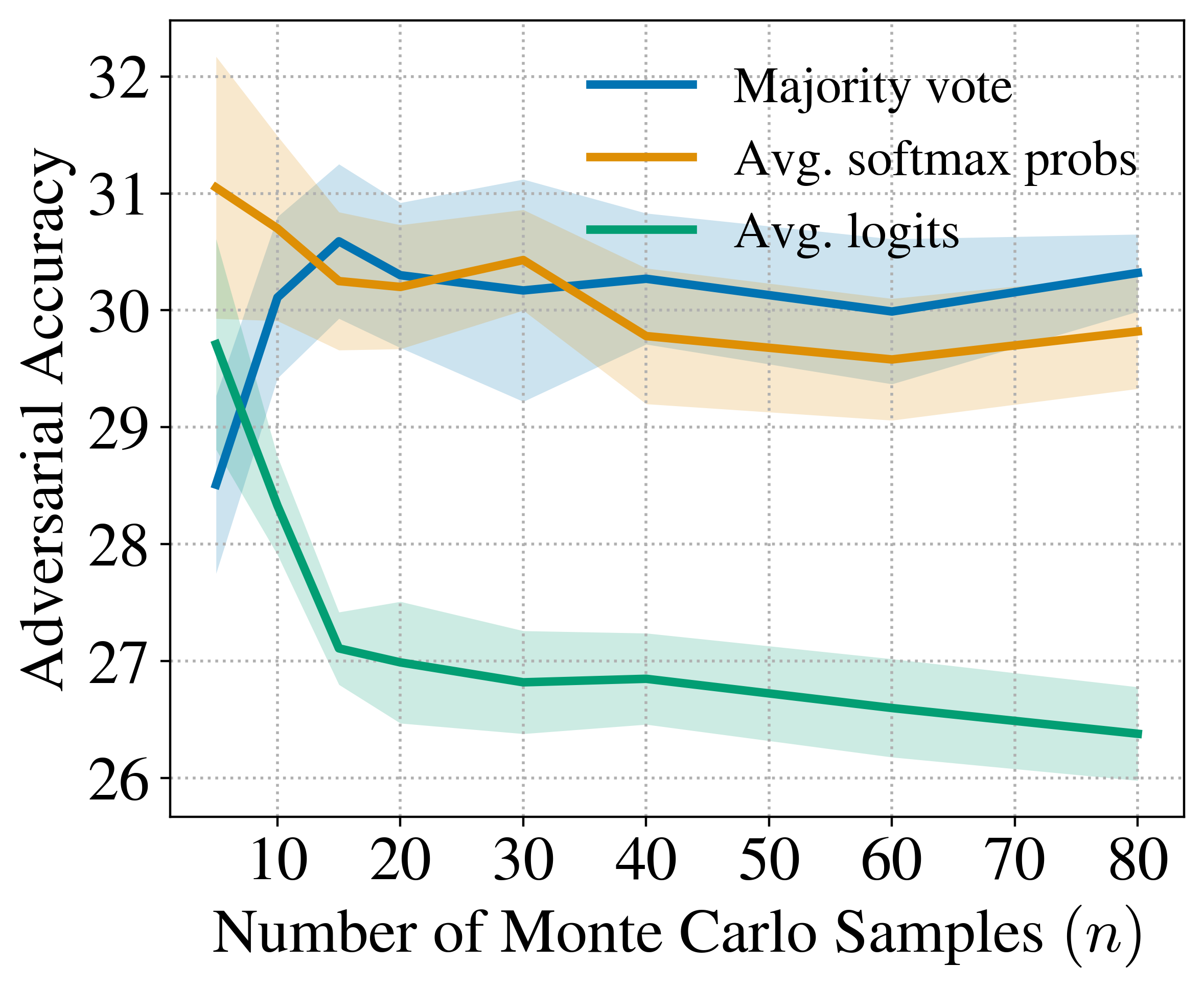}
    \caption{Adversarial accuracy ($\epsilon=16/255$) of our best \rt model computed with three decision rules for obtaining the final prediction from the $n$ output samples. The rules are majority vote (red), average softmax probability (blue), and average logits (green). The shaded areas represent the 95\% confidence interval for each decision rule.}
    \label{fig:adv_rule}
\end{figure}

\figref{fig:clean_rule} and \figref{fig:adv_rule} compare three different decision rules that aggregate the $n$ outputs of the \rt model to produce the final prediction $\hat{y}(x)$ given an input $x$.
We choose the average softmax probability rule for all of our \rt models because it provides a good trade-off between the clean accuracy and the robustness.
Majority vote has poor clean accuracy, and the average logits have poor robustness.

\subsection{Importance of the Transformation Groups} \label{ap:sec:rank}

\begin{table}[t]
\small
\centering
\caption{\rt's performance when only one of the transformation groups is applied. The attack is Linear+Adam+SGM with 200 steps and $n=20$.}
\label{tab:tf_group_used}
\begin{tabular}{@{}lrr@{}}
\toprule
Used Transformations & Clean Acc.       & Adv. Acc.        \\ \midrule
Noise injection         & $80.93 \pm 0.44$ & $\mathbf{8.35 \pm 0.20}$ \\
Blur filter             & $97.32 \pm 0.20$ & $0.00 \pm 0.00$ \\
Color space             & $94.40 \pm 0.53$ & $0.00 \pm 0.00$ \\
Edge detection          & $97.64 \pm 0.09$ & $0.00 \pm 0.00$ \\
Lossy compression       & $83.56 \pm 0.66$ & $3.56 \pm 0.26$ \\
Geometric transforms    & $88.42 \pm 0.28$ & $0.83 \pm 0.21$ \\
Stylization             & $\mathbf{98.31 \pm 0.09}$ & $0.00 \pm 0.00$ \\ \bottomrule
\end{tabular}
\end{table}

Choosing the best set of transformation types to use is a computationally expensive problem.
There are many more transformations that can be applied outside of the 33 types we choose, and the number of possible combinations grows exponentially.
BO gives us an approximate solution but is by no means perfect.
Here, we take a step further to understand the importance of each transformation group.
Table~\ref{tab:tf_group_used} gives an alternative way to gauge the contribution of each transformation group. 
According to this experiment, noise injection appears most robust followed by lossy compression and geometric transformations.
However, this result is not very informative as most of the groups have zero adversarial accuracy, and the rest are likely to also reduce to zero given more attack steps.
This result also surprisingly follows the commonly observed robustness-accuracy trade-off~\citep{tsipras_robustness_2019}.

\subsection{Number of Transformations} \label{ap:ssec:num_tf}

\begin{figure}[t!]
    \centering
    \includegraphics[width=0.4\textwidth]{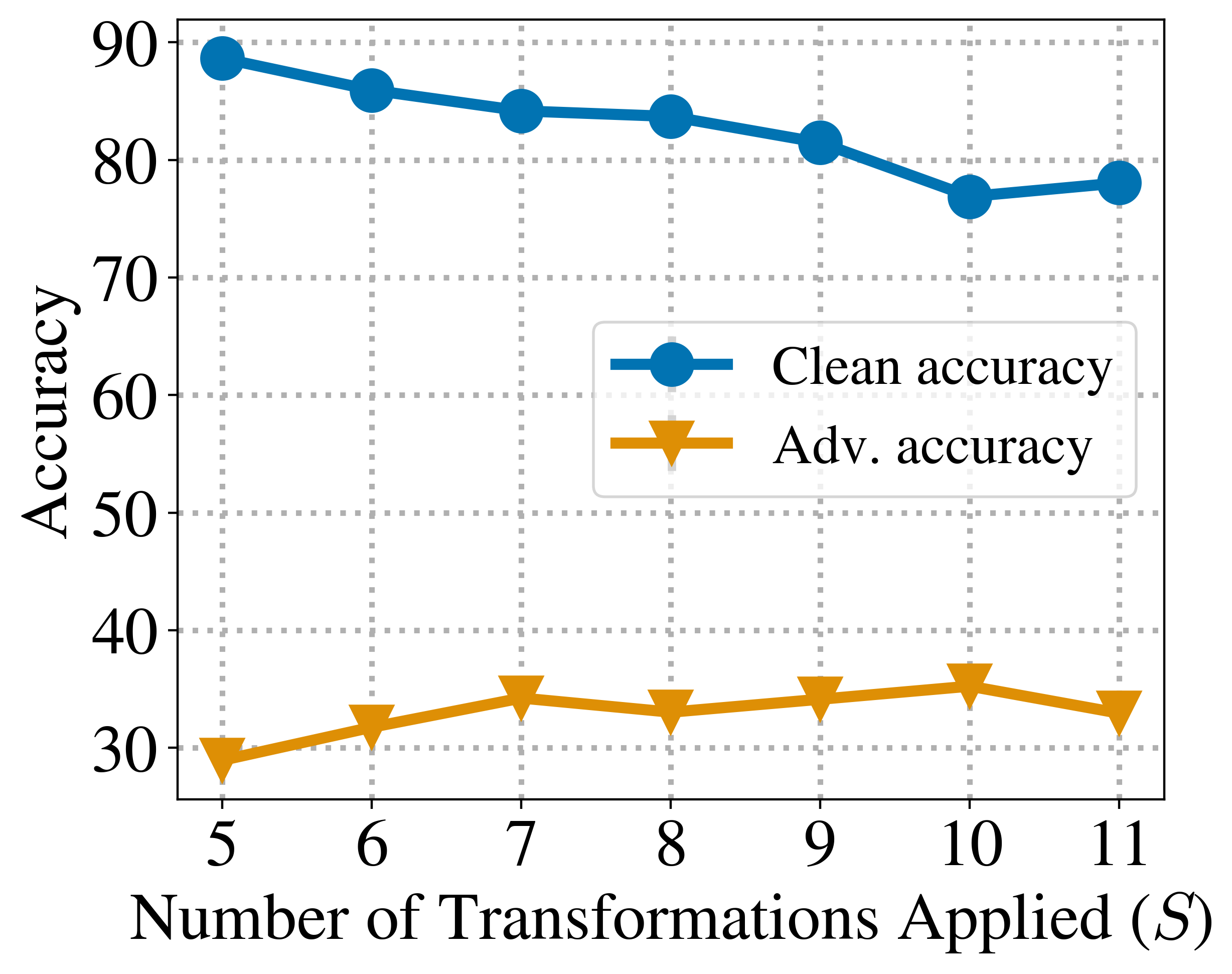}
    \captionof{figure}{Adversarial accuracy of \rt models obtained after running Algorithm~\ref{alg:bo} for different values of $S$ on CIFAR-10}
    \label{fig:num_tf}
\end{figure}

We test the effect of the transform permutation size $S$ on the clean and the robust accuracy of \rt models (\figref{fig:num_tf}).
We run Bayesian optimization experiments for different values of $S$ using all 33 transformation types, and all of the models are trained using the same procedure.
\figref{fig:num_tf} shows that generally more transformations (larger $S$) increase robustness but lower accuracy on benign samples.

\end{document}